\documentclass[11pt,preprint,a4paper,twocolumn]{elsarticle}
\usepackage[margin=0.8in]{geometry}
\usepackage{hyperref}
\biboptions{sort&compress}
\usepackage{bm}
\usepackage{gensymb}

\usepackage{amsmath}
\usepackage{color}

\usepackage{mathtools}

\usepackage{graphicx} 
\usepackage{amsfonts}
\usepackage{float}
\usepackage{multicol}
\usepackage[normalem]{ulem}

\bibpunct{[}{]}{,}{s}{,}{,}

\usepackage{array}   
\newcolumntype{P}[1]{>{\centering\arraybackslash}p{#1}}

\newcommand{\me}{\mathrm{e}}

\usepackage{soul}
\usepackage{placeins}
\usepackage{caption}
\begin{document}
\onecolumn
\begin{frontmatter}
\title{Quasi-Chemical Theory for Anion Hydration and Specific Ion 
Effects: Cl$^-$(aq) \emph{vs.} F$^-$(aq) }
\date{}
\author{A. Muralidharan$^a$, L. R. Pratt$^a$, M. I. Chaudhari$^b$, S. B. Rempe$^b$ $^\dagger$} 
\address{amuralid@tulane.edu, lpratt@tulane.edu, michaud@sandia.gov,
slrempe@sandia.gov}
\address{%
$^a$Department of Chemical and Biomolecular Engineering, Tulane University, New Orleans, LA 70118, U.S.A.\\%
$^b$Department of Nanobiology, Sandia National Laboratories, Albuquerque, NM 87185, U.S.A. \\%
$^\dagger$Corresponding author: 505-845-0253
}

\begin{abstract}
Anion hydration is complicated by H-bond donation between neighboring
water molecules in addition to H-bond donation to the anion. This
situation can lead to competing structures for chemically simple
clusters like (H$_2$O)$_n$Cl$^-$ and to anharmonic vibrational motions.
Quasi-chemical theory builds from electronic structure treatment of
isolated ion-water clusters, partitions the hydration free energy into
inner-shell and outer-shell contributions, and provides a general
statistical mechanical framework to study complications of anion
hydration. The present study exploits dynamics calculations on isolated
(H$_2$O)$_n$Cl$^-$ clusters to account for anharmonicity, utilizing ADMP
(atom-centered basis sets and density-matrix propagation) tools.
Comparing singly hydrated F$^-$ and Cl$^-$ clusters, classic OH-bond
donation to the anion occurs for F$^-$, while Cl$^-$ clusters exhibit
more flexible but dipole-dominated interactions between ligand and ion.
The predicted Cl$^-$ -- F$^-$ hydration free energy difference agrees
well with experiment, a significant theoretical step for addressing
issues like Hofmeister ranking and selectivity in ion channels.

\end{abstract}

\end{frontmatter}

\twocolumn
\section{Introduction}
Here, we build a fundamental molecular statistical mechanical theory for
hydration of chloride ion (Cl$^-$). Theory of the type targeted here,
specifically quasi-chemical theory (QCT), permits transparent comparison
of Cl$^-$ to analogous cases such as fluoride hydration (F$^-$).
Comparisons of that sort underpin specific ion effects in areas of broad
current interest, including the Hofmeister
ranking\cite{Jungwirth:2006gu,Kalcher:2009js,%
Zhang:2010gr,Kunz:2010fy,dpars11,Pollard:2016ei} of ions and the
mechanism of specific ion permeation in membrane ion channels. An example of membrane ion channels is the CLC family, which regulate membrane transport of
Cl$^-$.\cite{SPERELAKIS2012121} Though we comment further below on such
long-term goals, we focus primarily on extending QCT to hydrated anions.
One  motivation for comparisons such as Cl$^-$ with F$^-$ is that they
permit rigorous experimental thermodynamic testing of single-ion free
energies without discussion of potentials of the
phase,\cite{you2014comparison,pratt1992contact} or 
surface potentials.\cite{Leung:2009dx,lyklema2017interfacial,doyle2019importance}

Adequate molecular simulation is necessary for compelling molecular
statistical thermodynamic theory of liquid solutions, and extensive
simulation studies have been carried
out.\cite{Jungwirth:2002et,Jungwirth:2002eu,Tongraar:2003bd,%
Heuft:2005jt,Heuft:2005kxa,%
Heuft:2003iva,Ho:2009bga,Raugei:2002gf,%
cabarcos1999microscopic,bankura2013hydration} Available computational
work has nicely delineated distinctions in hydration structure of anions
compared to simple metal ions, and distinctions between different halide
ions.\cite{Perera:1994ii,Sremaniak:1994wm,Herce:2005dp,%
Eggimann:2007cp}
In this crowded landscape of computational work, molecular-scale
statistical mechanical theory for these systems has been meager, with
striking exceptions.\cite{Bostrom:2005km}

Quasi-chemical theory (QCT), in the present
incarnation based on clusters,\cite{asthagiri2010ion,%
rogers2012:structural} provides a natural starting point for the theory
sought.\cite{Rogers:2010gh,Pollard:2016ei}
QCT\cite{pratt1999quasi,rempe2000hydration,paliwal2006analysis,%
beck2006potential,shah2007balancing,asthagiri2010ion,%
rogers2012:structural} naturally divides
the free energy determination into inner-shell and outer-shell
considerations, as is apparent from Eq.~\eqref{eq:qct} below. In this
approach, the necessary contributions are physically
meaningful,\cite{Anonymous:2009dc} and can then be evaluated to the
desired accuracy --- even judiciously using experimental results for
natural intermediate results.\cite{muralidharan2018quasi}  QCT,
considered broadly over nearly two decades, has been applied
successfully to a wide range of systems, including the hard sphere
fluid,\cite{pratt2003self} liquid
water,\cite{shah2007balancing,Weber:2011hd,Anonymous:2009dc}, 
small molecule solvation\cite{sabo2006:H2,sabo2008hydrogen,%
clawson2010:H2,jiao2012combined,jiao2011co2,Chaudhari:2015jq}
also involving ion channels\cite{varma2007tuning,varma2011design,%
rogers2011:probing,chaudhari2018strontium}
and other macromolecules,\cite{varma2008valinomycin,Chaudhari:2014ga,%
stevens2016:carboxylate} cations in water and
non-aqueous solvents,\cite{rempe2000hydration,rempe2001hydration,%
AsthagiriD:QuasB2,rempe2004inner,asthagiri2004hydration,%
varma2008structural,jiao2010:Ni,Sabo:2013gs,chaudhari2015:Ba,%
chaudhari2016:carbonate,chaudhari2018utility} and biomolecule
hydration.\cite{weber2012regularizing,Tomar:2013dt} 

The QCT approach provides a complete statistical
mechanical framework for evaluating solvation free energies of
ions by building up from carefully defined isolated 
ion-solvent clusters.\cite{asthagiri2010ion} Experimental
measurements addressing the energetics of such
clusters, based on mass 
spectroscopy,\cite{arshadi1970hydration,keesee1980properties,%
hiraoka1988solvation} have existed for some 
decades. For hydrous clusters, a
compilation of those experimental values are available from Tissandier,
{\emph {et al.}}\cite{tissandier1998proton}

\begin{figure}[!htbp]
\centering \boxed{ \includegraphics[width=0.45\textwidth]{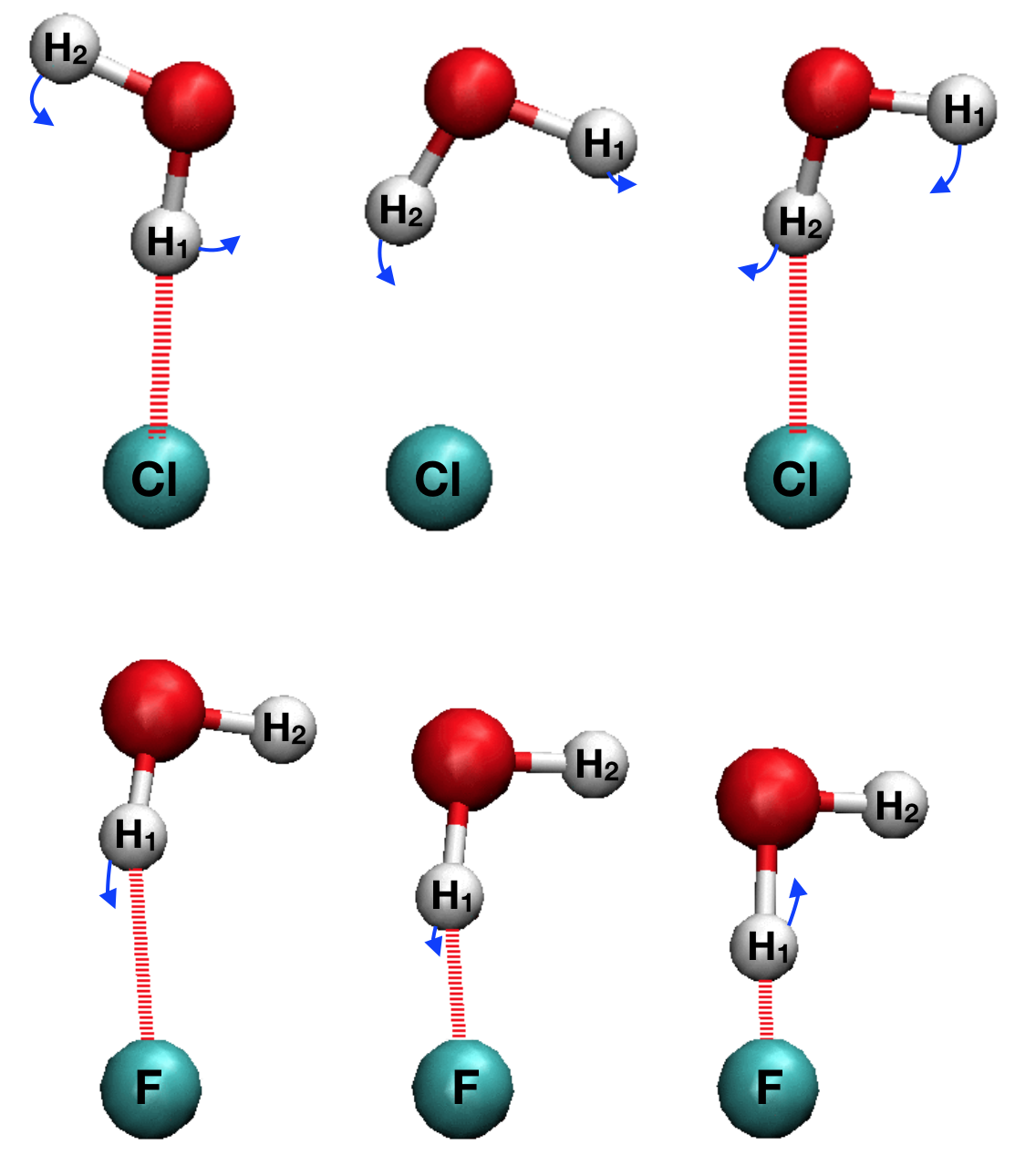}}
\caption{Sequence of structures over about 0.1~ps of the ADMP classical
dynamics calculations.   An animated display is available at  (\url{https://doi.org/10.6084/m9.figshare.8066270})
and discussed in Section \ref{sec:anim}. The qualitative impression
is of $\left(\mathrm{H}_2\mathrm{O}\right)\mathrm{Cl}^-$ (upper) rocking
through a dipole-dominated configuration, in contrast to (lower) OHF
stretching vibration for the
$\left(\mathrm{H}_2\mathrm{O}\right)\mathrm{F}^-$ cluster. }
\label{fig:snapshots} 
\end{figure}

QCT has been applied less to the challenging cases of anion hydration
than to metal cations, comparatively. A primary difficulty is hydrogen
bonding between water molecules of clusters such as
(H$_2$O)$_n\mathrm{Cl}^-$, leading to structures that are less
susceptible to simple approximation in the implementation of QCT. Recent
efforts have been directed toward addressing those initial
approximations in the application of QCT. Some of the refinements
include quantification of anharmonic
effects\cite{sabo2008hydrogen,jiao2011co2,rogers2011:probing,%
muralidharan2018quasi} on free energies of solute-water clusters and the
sufficiency of the polarizable continuum
model\cite{Tomasi:2005tc,sabo2008hydrogen,chaudhari2015:Ba} (PCM) for
the hydration free energy of those clusters. 



Anticipating results below (Section \ref{sec:cluster-energetics}),  a
harmonic approximation to the potential energy surface of
(H$_2$O)$_n\mathrm{Cl}^-$ clusters is satisfactory for cluster formation
enthalpies, but not for free energies.  To investigate such issues, we
use a molecular dynamics approach that explicitly includes electron
coordinates, as we now discuss. 

Classical trajectory calculations that include electron coordinates fall
into two major categories: Born-Oppenheimer (BO) molecular dynamics (MD)
and extended Lagrangian (EL) dynamics. In the BO approach, the atomic
dynamics develop on a potential energy surface obtained from an
optimized-in-detail electronic structure calculation. In EL approaches
such as Car-Parrinello molecular dynamics (CPMD),\cite{CPMD} electron
coordinates are introduced as  fictitious dynamic variables that couple
electronic and atomic motion. The dynamics should keep the fictitious
electronic degrees of freedom near the ground state BO potential energy 
surface, yielding
sufficiently accurate forces for the atomic trajectory.  

The \emph{atom-centered basis sets and density matrix propagation}
(ADMP)\cite{schlegel2002ab} tool is an extended Lagrangian approach for
molecular dynamics. In contrast with CPMD, ADMP emphasizes finite
non-periodic systems, and thus is convenient for the 
isolated clusters of interest here 
(Figure \ref{fig:snapshots}). ADMP
has been shown to provide similar functionality to BO molecular
dynamics.\cite{schlegel2002ab}

\begin{figure}[tbp]
\centering \includegraphics[width=0.49\textwidth]{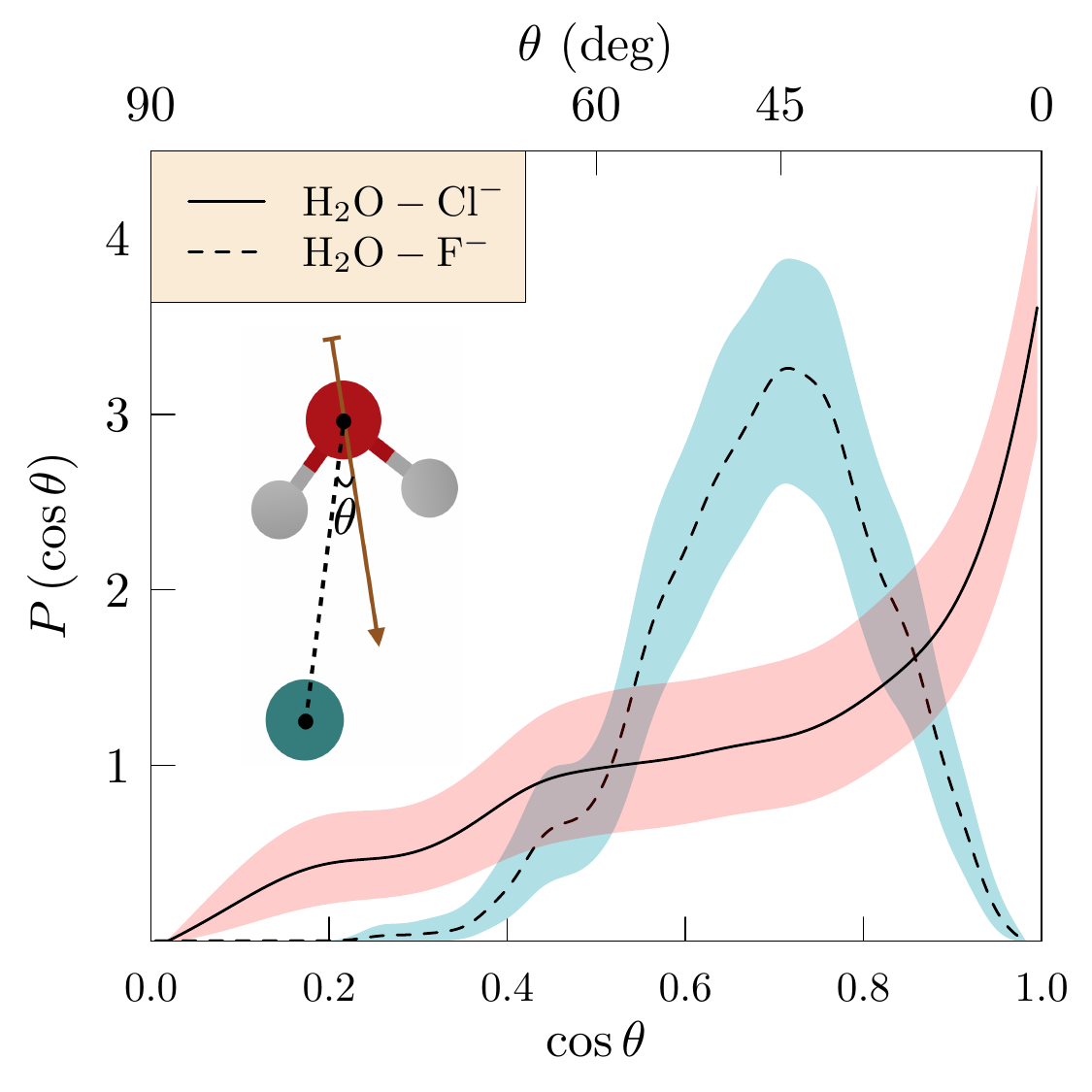}
\caption{Distribution of $\cos \theta$, with $\theta$ the angle between
water dipole moment and the OX$^-$ vector, observed over the last 5~ps
of a 10~ps ADMP trajectory at $T$=300~K. For
$\mathrm{H_2O}- \mathrm{F}^-$, the distribution peaks at 45\degree~
with an OH bond oriented toward the F$^-$, as shown in Figure
\ref{fig:snapshots}. In contrast,  the
$\theta$ distribution spans from 0\degree ~to 90\degree, ~with
maximum near 0\degree, for  $\mathrm{H_2O}- \mathrm{Cl}^-$. Rocking through dipole-dominated
configurations allows the water molecule to offer alternate H atoms for
coordination with the ion.  The filled bands indicate  approximate 
99\% confidence intervals, estimated on the basis of a resampling
(bootstrap) procedure.} 
\label{fig:angle} 
\end{figure}

\subsection{Perspective from interest in Cl$^-$ and F$^-$ ion channels}
The regulation of ion concentration is essential for several
physiological functions in cells.\cite{alvarez2009physiology} For anions
such as Cl$^-$, this control is achieved through membrane channels and transport
proteins from the CLC family, which facilitate the passage of Cl$^-$
through electro-chemical potential gradients.\cite{SPERELAKIS2012121}
Numerous efforts\cite{park2017structure,jentsch2002molecular,%
robertson2010design,accardi2010clc} have been directed toward
deciphering the structures of those proteins and understanding their
functional characteristics, such as conductance, gating, and ion
selectivity. 

While CLC proteins select for Cl$^-$, other channel proteins
discriminate against Cl$^-$. An interesting example is FLUC, a family of
fluoride-specific ion channels with dual-topology
architecture.\cite{stockbridge2015crystal} These channels display an
astonishingly high selectivity of $10^{4}$ for F$^{-}$ over Cl$^-$
despite close similarity in size and identical charge of the ions. FLUC
channels provide a ladder of hydrogen-bond donating residues that
apparently create a polar track for F$^-$. How that polar track leads to
the unusually high level of discrimination between F$^-$ and Cl$^-$
remains an open question. Understanding such mechanisms for selectivity
in ion transport has been a primary target for many recent modeling and
simulation studies, emphasizing K$^+$/Na$^+$ selectivity of potassium
ion 
channels.\cite{Sansom2002,varma2007tuning,Corry:2007,%
bostick2007selectivity,varma2008valinomycin,Fowler2008,bostick2009statistical,roux2011ion,%
varma2011design,noskov2011importance,dixit2011thermodynamics,Kim2011,furini2011,%
Andersen393,Alam397,Nimigean405,roux2016,deGroot2018}  
But computational studies of anion transport mechanisms  
demonstrated by  CLC channels\cite{gervasio2006exploring,%
ko2010chloride,kuang2008transpath,yin2004ion,smith2011charge,%
chen2016free,cheng2012molecular} that address comparison to 
alternatives such as FLUC are less mature.

To address ion selectivity between different media, a rigorous treatment
would require comparison of the total free energy of ion solvation
between those environments. However, ion channels are thought to
function primarily by manipulation of the local environment of the ion.
That manipulation may arise from constraints on local structure imposed
by the structural and chemical properties of the surrounding environment
or the binding sites on the channel itself.\cite{varma2007tuning,%
Jordan:2007Notable,Corry:2007,bostick2007selectivity,%
varma2008valinomycin,varma2008structural,bostick2009statistical,%
varma2011design,rogers2011:probing,rogers2012:structural,%
stevens2016:carboxylate,chaudhari2018strontium} 
Furthermore, contributions to solvation free energy from long-ranged
electrostatic interactions between an ion and the distant environment
might cancel between ions of the same charge. Therefore, analysis of
selectivity might not require a detailed description of the long-ranged
interactions in the molecular theory. Instead, rigorous treatment of
local interactions are likely important for accurate treatment of ion
selectivity. 

Here, we compare the hydration of Cl$^-$ with F$^-$, focusing on
rigorous evaluation of local interactions. Broader issues such as the
Hofmeister ranking of ions and selectivity in membrane channels are
reserved for the future. In the next two sections we lay out the theory
and results, followed by a concluding discussion. The details of the
numerical implementation are specified in Section~\ref{methods}.  

\section{Theory: Step-wise evaluation of the 
isolated cluster free energy} \label{theory:scheme} 

QCT formulates the net hydration free energy as
\begin{multline}
\mu^{\mathrm{(ex)}}_{\mathrm{X}^{-}} = -RT\ln K^{(0)}_{n}\rho_{\mathrm{H_2O}}{}^{n}
	+RT\ln p_{\mathrm{X}^{-}}(n) \\
	+\left(\mu^{\mathrm{(ex)}}_{\left(\mathrm{H}_2\mathrm{O}\right)_n\mathrm{X}^-} -n \mu^{\mathrm{(ex)}}_{\mathrm{H}_2\mathrm{O}}\right)~,
	\label{eq:qct}
\end{multline}
collecting the individual quasi-chemical
contributions.\cite{asthagiri2010ion,rogers2012:structural}
The first and last term represent the inner-shell and
outer-shell contributions defined based on assignment of a clustering
radius ($\lambda$). $p_{\mathrm{X}^{-}}(n)$ then identifies the thermal
probability of observing $n$ waters within that prescribed $\lambda$. 
The several features of this expression will be identified in the
following discussion.

To that end, consider the chemical process 
\begin{eqnarray} n\mathrm{H_2O} + \mathrm{X}{}^- 
\xrightleftharpoons[]{K^{(0)}_n} 
\mathrm{\left(H_2O\right)}_n\mathrm{X}{}^-~
\label{eq:chemprocess}
\end{eqnarray}
for formation of the isolated cluster, as in an ideal gas phase (X = Cl here). We seek
the free energy (inner-shell) contribution $-RT \ln K_{n}^{(0)}\rho_{\mathrm{H_2O}}{}^{n}$, with
\begin{eqnarray} 
K_n ^{(0)} &=& \frac{\mathcal{Q}\left(\gamma_n\sigma \right)/ n! }
{ \mathcal{Q}(\sigma) \left[\mathcal{Q}(\gamma)/V \right] ^n}~,
\label{Eq:SI}
\end{eqnarray}
a classic aspect of statistical
thermodynamics.\cite{beck2006potential,muralidharan2018quasi} Here $V$
is the system volume, $\sigma$ and $\gamma$ represent the solute
(Cl$^-$) and solvent (H$_2$O) molecules, respectively, and the
$\mathcal{Q}\left(\gamma_n\sigma \right)$ are configurational integrals
associated with single molecule/cluster canonical partition
functions.\cite{muralidharan2018quasi}

Our scheme for evaluating $K_{n}{}^{(0)}$ proceeds step-wise in $n$
according to 
\begin{multline} 
\frac{K_{n}{}^{(0)}}
{K_{n-1}{}^{(0)}K_{1}{}^{(0)}} = \\
\frac{ 
\mathcal{Q}\left(\gamma_n\sigma \right)/\mathcal{Q}(\sigma) 
}{
    n \left \lbrack 
    \mathcal{Q}\left(\gamma_{n-1}\sigma \right)/\mathcal{Q}(\sigma) 
\right \rbrack 
\left \lbrack
\mathcal{Q}\left(\gamma\sigma \right) 
/ \mathcal{Q}(\sigma) \right \rbrack 
}~.
\label{Eq:step_fe_1} 
\end{multline}
The numerator on the right of Eq.~\eqref{Eq:step_fe_1} involves
integration carried over canonical distributions of configurations  of
$n$ solvent molecules, while the denominator treats configurations of
$n-1$ and 1, respectively, treated independently. Our scheme is then
based on evaluation of  
\begin{multline}
\Delta U = E\left(\gamma_n\sigma \right) - E\left(\gamma_{n-1}\sigma \right)\\ - E\left(\gamma\sigma \right) + E(\sigma)~,
\label{Eq:step_fe_2} 
\end{multline} 
so that
\begin{eqnarray} 
n K_{n}{}^{(0)} = 
\frac{K_{1}{}^{(0)}K_{n-1}{}^{(0)}}{\left \langle e^{\beta \Delta U} \right \rangle _n}~.
\label{Eq:final} 
\end{eqnarray}
The brackets, $\langle \ldots \rangle_n$, indicate the thermal average
utilizing configurations from the canonical simulation stream for the
$\gamma_n\sigma$ cluster. 

For $n=1$, Eq.~\eqref{Eq:final} reduces to the trivial case of
$K_{0}{}^{(0)} = 1$. In the evaluation of $K_{n}{}^{(0)}$ for $n\geq2$,
the value of $K_{1}{}^{(0)}$ can be supplied from
experiment\cite{tissandier1998proton} or theory. This term describes the
interaction between Cl$^-$ and one water molecule. We note that
for (H$_2$O)Cl$^-$, a harmonic approximation gives a sufficiently good
estimate for the free energy, but not for bigger cluster sizes (Figure
\ref{fig:cluster_formation_FE}). Carrying out subsequent steps in this
scheme then addresses the issues that make anion hydration more
challenging; that is, the sampling of cluster configurations,
\cite{muralidharan2018quasi} including competing H-bonding interactions
between neighboring water molecules in those clusters.

Here, we take advantage of the QCT partitioning to evaluate the
inner-shell contribution based on accurate quantum mechanical treatment
of dynamics (using ADMP) in ion-water clusters. The evaluation of
$\left(
\mu^{\mathrm{(ex)}}_{\mathrm{(H_2O)}_n\mathrm{Cl}^-}-n\mu^{\mathrm{(ex)
}}_{\mathrm{H_2O}}
\right)$ utilizes the same trajectory, but approximates the outer-shell
waters using the polarizable continuum model\cite{Tomasi:2005tc} (PCM),
as described in section \ref{methods}. Experimental comparisons (section
\ref{fig:Cl_hydration_FE}) can then indicate the sufficiency of such an
approximation.

\begin{figure*}[tbp]
\begin{center}
\includegraphics[width=0.49\textwidth]{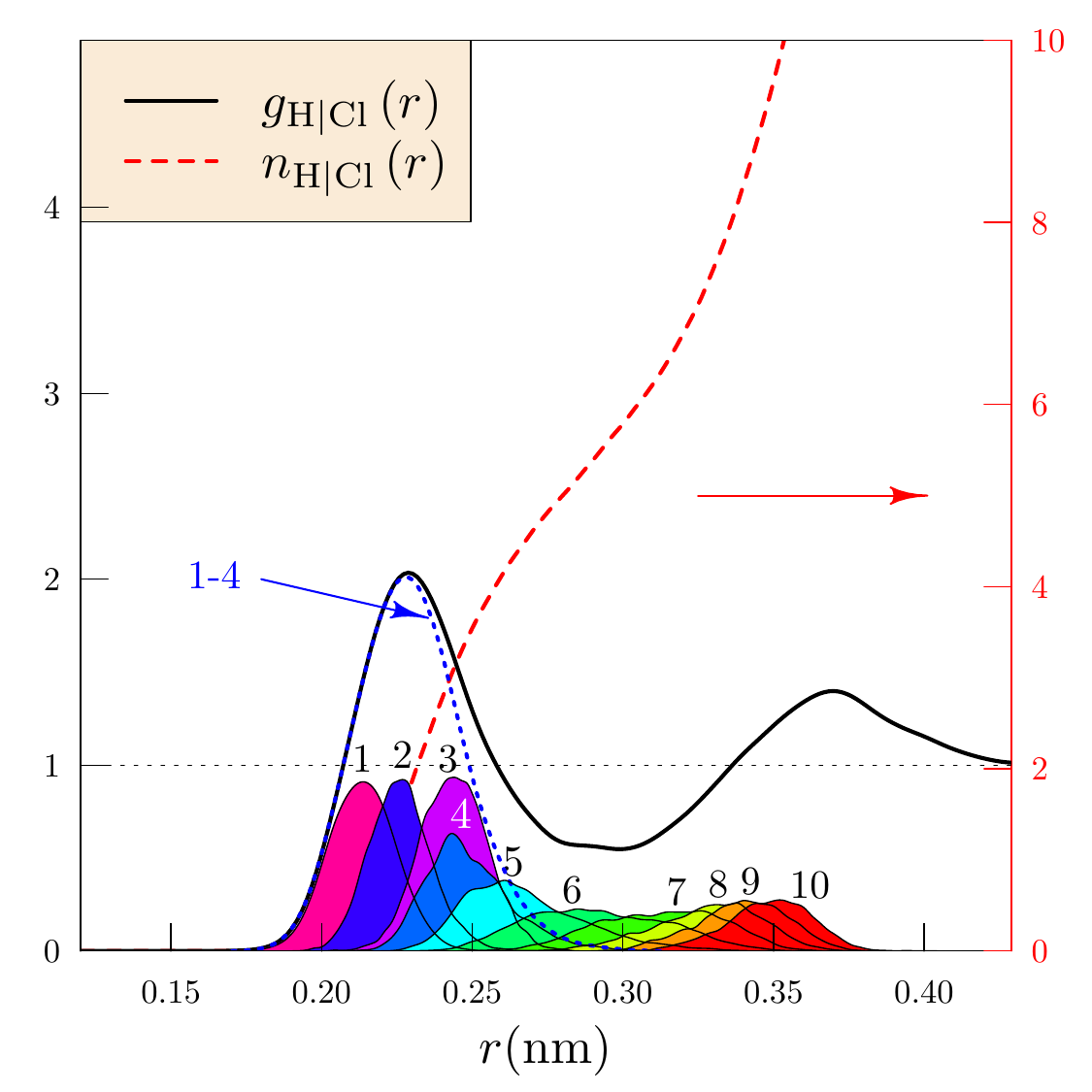}
\hfill
\includegraphics[width=0.49\textwidth]{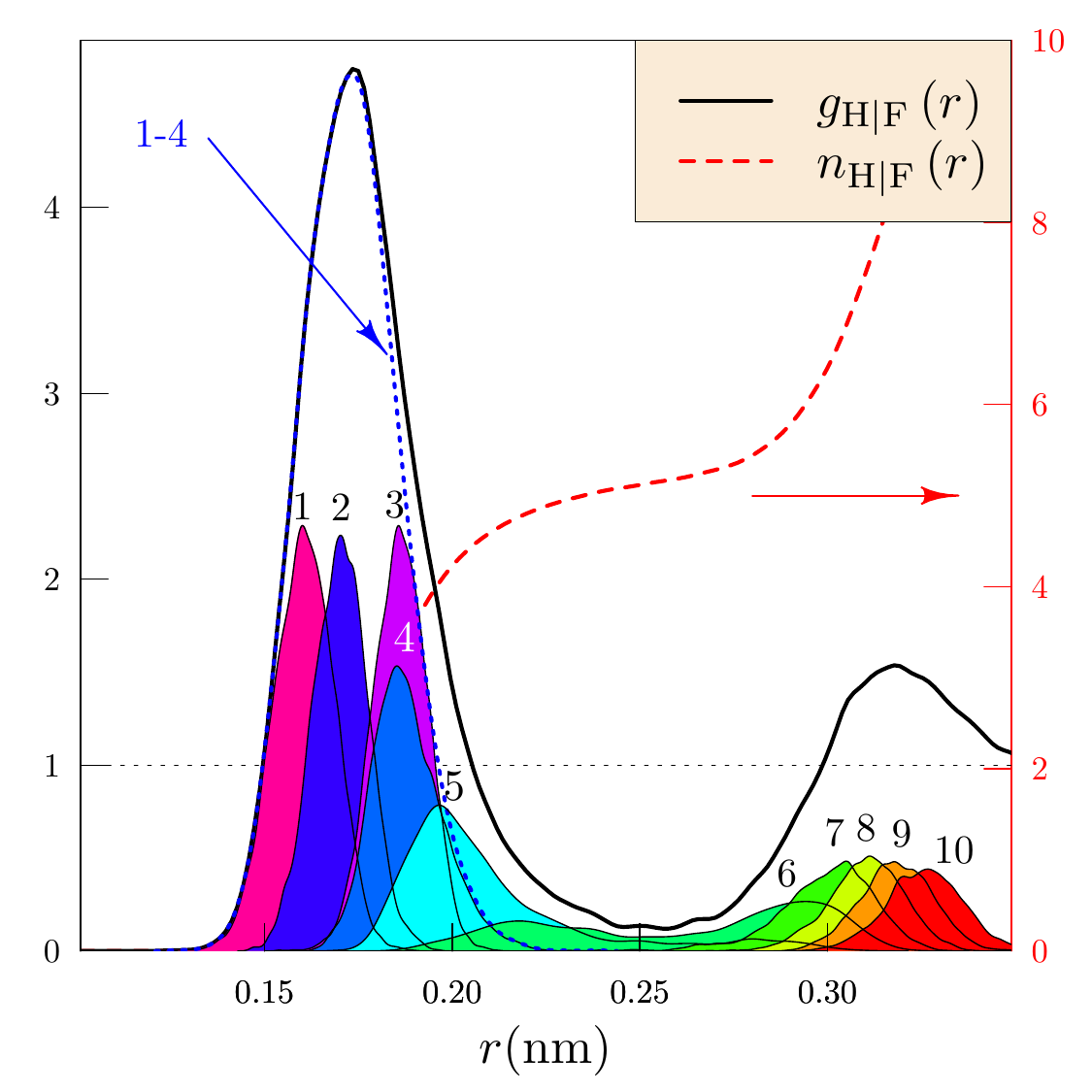}
\caption{Radial distributions and running coordination number  of H
atoms relative to X$^-$ from AIMD simulations of X$^-$ in bulk water
phase (X = Cl or F). The integer-labeled distributions are the
neighborship-ordered contributions of the $n^\text{th}$ nearest H atom. A
choice $\lambda_{\mathrm{Cl}^-} \leq 0.26$ nm and
$\lambda_{\mathrm{F}^-} \leq 0.20$ nm excludes split-shell clusters.
Compared to Cl$^-$(aq), the radial distribution of F$^-$(aq) displays a
larger maximum and a nearly zero first minimum, leading to a distinct
plateau of the running coordination number.}
\label{fig:Neighborship_H} 
\end{center}
\end{figure*}

\section{Results}
\label{sec:results}
\subsection{Solution structure}\label{theory:formulation}
Neighborship is a fundamental concept in
QCT.\cite{varma2007tuning,asthagiri2010ion,Sabo:2013gs,muralidharan2018quasi,ACR2019} 
The H-atom of water is the closest neighbor to the anions, F$^-$ and Cl$^-$. Proximity
distinctions between those cases are sharper when H atoms are utilized
rather than O atoms in the radial distributions. Distributions of water
H atoms relative to the ions (Figure~\ref{fig:Neighborship_H}) then
guide the assignment of the clustering radius, $\lambda$. Further, the
neighborship-ordered contributions help identify cluster sizes in which
waters are always in direct contact with the ion. Notice that the
distribution of the 6$^\text{th}$ nearest water around Cl$^-$ extends beyond the first
minimum of $g_{\mathrm{H\vert Cl}}\left(r\right)$. For instance, a
choice of $\lambda_{\mathrm{Cl}^-} \leq 0.26$ nm eliminates such
split-shell
clusters.\cite{varma2007tuning,Sabo:2013gs,chaudhari2017quasi} The thermal probability
($p_{\mathrm{Cl}^{-}}(n)$), contributing to the second term of Eq.
\eqref{eq:qct}, is then evaluated based on this constraint.

Compared to Cl$^-$(aq), the radial distribution of F$^-$(aq) displays a
larger maximum  and a nearly zero first minimum, leading to a flat
behavior of the running coordination number above the minimum
(Figure~\ref{fig:Neighborship_H}). For F$^-$(aq), this structure suggests tightly
bound inner-shell waters, with small deviations from orientations that
take the nearest H-atom away from the ion. By comparison, inner-shell
waters of Cl$^-$(aq) are more flexibly structured. The differences
highlighted here  also feature in the behavior of waters in the gas
phase clusters of $\left(\mathrm{H}_2\mathrm{O}\right)_n\mathrm{Cl}^-$
and $\left(\mathrm{H}_2\mathrm{O}\right)_n\mathrm{F}^-$ (Figures
\ref{fig:snapshots} and \ref{fig:angle}).

\subsection{Isolated cluster energetics: $\left(\mathrm{H}_2\mathrm{O}\right)_n\mathrm{Cl}^-$}
\label{sec:cluster-energetics}
\begin{figure}[tbp]
\centering \includegraphics[width=0.49\textwidth]{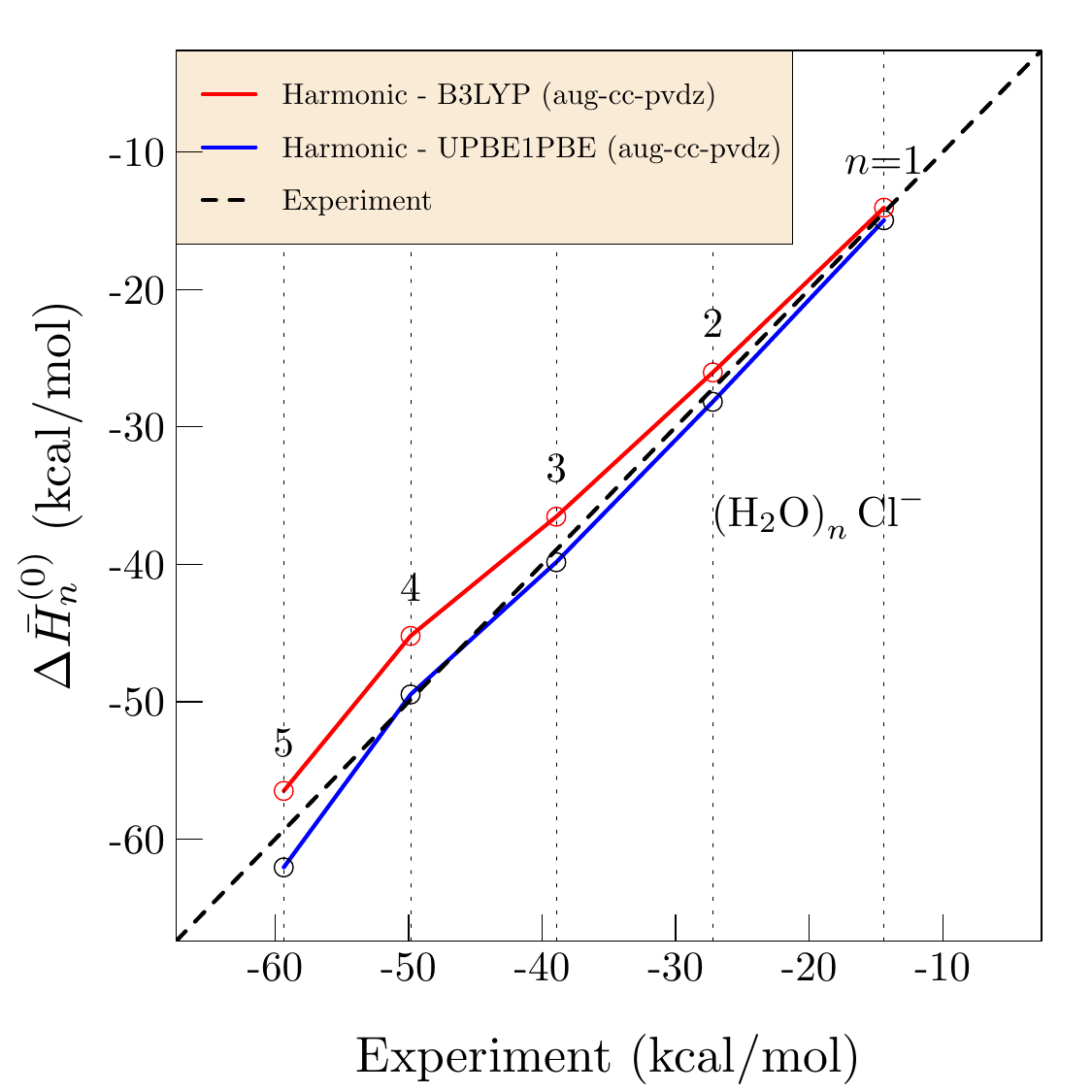}
\caption[Enthalpy of formation of
$\left(\mathrm{H}_2\mathrm{O}\right)_n\mathrm{Cl}^-$ clusters in the gas
phase]{Comparison of harmonic approximation with cluster
experiments\cite{tissandier1998proton} for $\Delta \bar{H}^{(0)}_{n}$,
the partial molar enthalpy of
$\left(\mathrm{H}_2\mathrm{O}\right)_n\mathrm{Cl}^-$ in the ideal gas
phase. Both functionals tested here satisfactorily follow the observed
trend. The UPBE1PBE functional, more accurate in this comparison, is
adopted as a basis for further QCT analysis of
$\left(\mathrm{H}_2\mathrm{O}\right)_n\mathrm{Cl}^-$.The 
experimental standard state is the ideal gas at $\left(T,p\right)$ =
(298~K,1~atm).}
\label{fig:Cl_cluster_formation_H} 
\end{figure}

To evaluate cluster energetics, a harmonic approximation was applied
initially to the lowest energy geometry-optimized structures (SI, Figure
S1). The enthalpies of $\left(\mathrm{H}_2\mathrm{O}\right)_n\mathrm{Cl}^-$
cluster calculated this way agree well with experiment
(Figure~\ref{fig:Cl_cluster_formation_H}). The UPBE1PBE functional with
the aug-cc-pvdz basis set showed the best agreement among the two tested
functionals. However, the free energy evaluated using the harmonic
approximation deviates from experiment, and that deviation increases with
$n$ (left panel, Figure~\ref{fig:cluster_formation_FE}). We conclude that
the entropy of the $\left(\mathrm{H}_2\mathrm{O}\right)_n\mathrm{Cl}^-$
cluster is not satisfactorily predicted in our harmonic approximation.

Notice further  that the harmonic approximation with the UPBE1PBE
functional gives an accurate estimate of free energy for
$\left(\mathrm{H}_2\mathrm{O}\right)\mathrm{Cl}^-$. Therefore, we expect
the discrepancy in the free energy of
$\left(\mathrm{H}_2\mathrm{O}\right)_2\mathrm{Cl}^-$ to emerge from
interactions between water molecules, roughly with an average strength
of an H-bond. Hence, we perform molecular dynamics of the cluster with
ADMP for better sampling of relative configurations of water in the
cluster. Those configurations are then utilized in the 
scheme  above (Section \ref{theory:scheme}) for evaluation of 
the free energy.
The free energies evaluated this way are in excellent agreement with experimental gas
phase cluster data (Figure~\ref{fig:cluster_formation_FE}).  
Our scheme (Section \ref{theory:scheme}) is highly
effective here because it is applied to the isolated cluster
and interactions with only a single surface water molecule are manipulated.

\subsection{Comparison with $\left(\mathrm{H}_2\mathrm{O}\right)_n\mathrm{F}^-$}
\label{sec:anim}
\begin{figure*}[tbp]
\begin{center}
\includegraphics[width=0.49\textwidth]{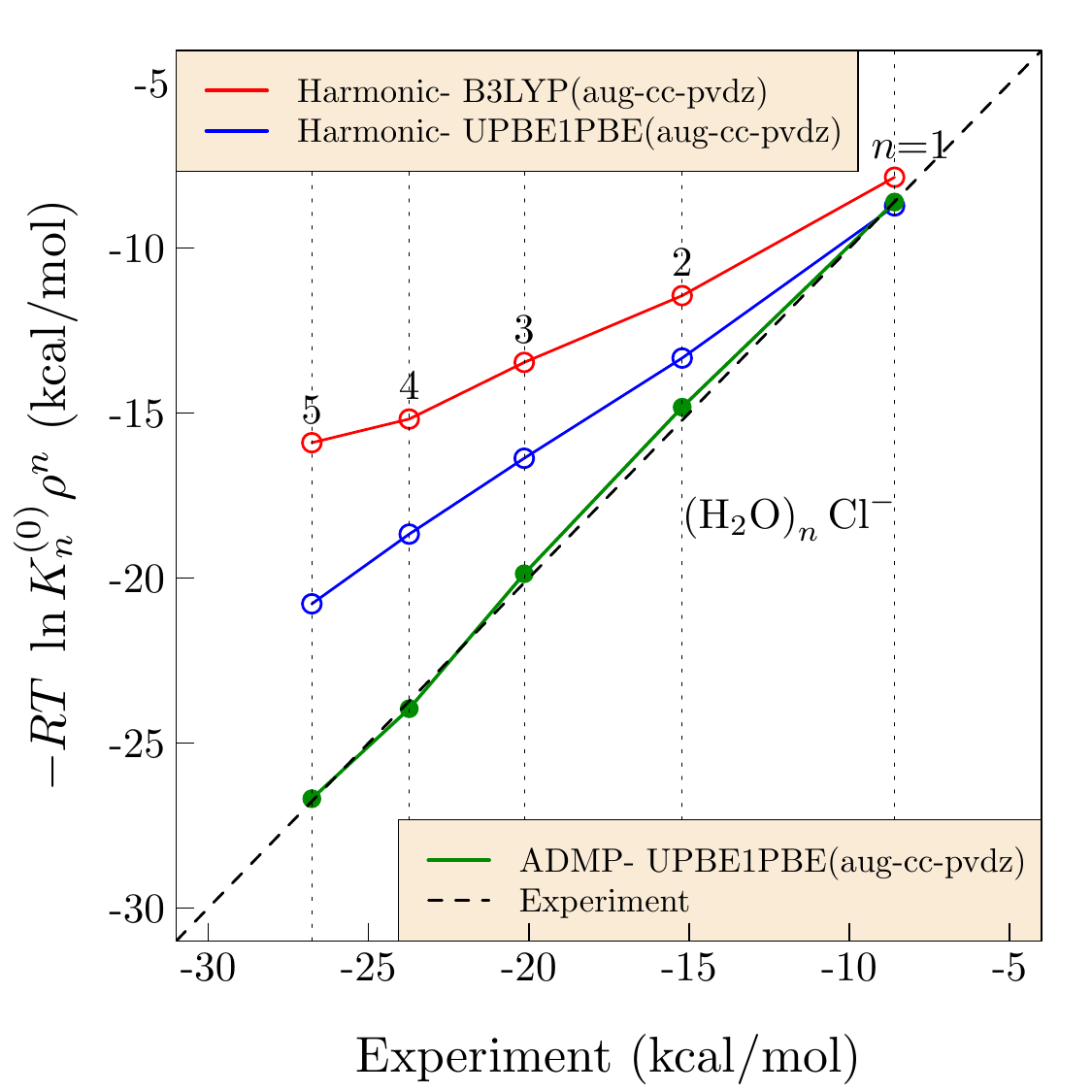}
\hfill
\includegraphics[width=0.49\textwidth]{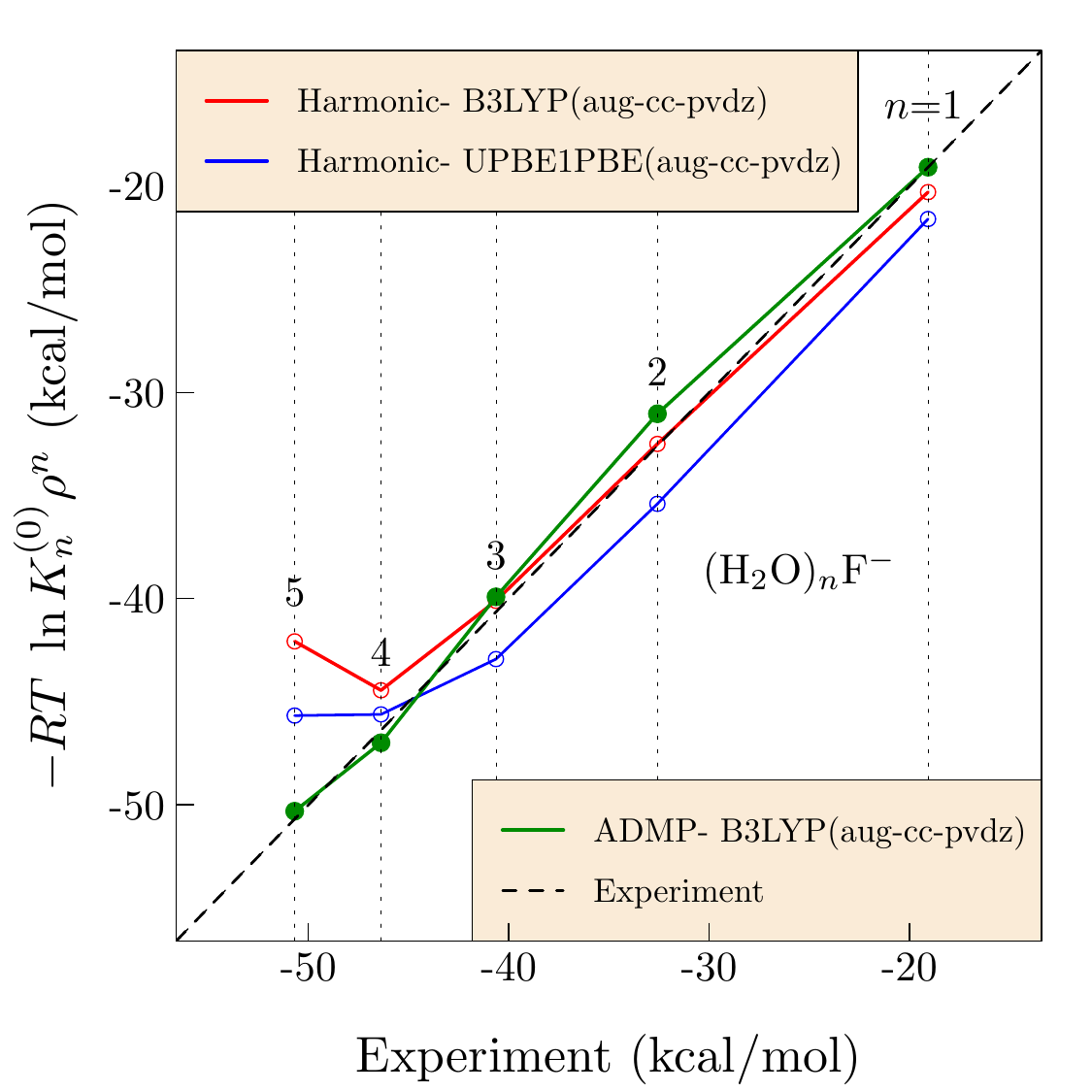}
\caption[Free energy of (H$_2$O)$_n\mathrm{X}^-$ clusters in gas
phase]{Evaluation of the free energy $-RT\ln K^{(0)}_{n}\rho^n$ for 
(H$_2$O)$_n\mathrm{X}^-$ clusters, for $n\leq 5$ (X = Cl or F). The
results make explicit the limitation of a harmonic approximation for
(H$_2$O)$_n\mathrm{Cl}^-$. The fuller analysis of the dynamics of the
cluster through ADMP yields excellent comparison with cluster
experiments.\cite{tissandier1998proton} The 
experimental standard state is the ideal gas at $\left(T,p\right)$ =
(298~K,1~atm).}
\label{fig:cluster_formation_FE} 
\end{center}
\end{figure*}

In light of the successful treatment of chloride clusters with ADMP, we
next take up the case of fluoride. For comparison, the same steps are
followed as before (Section \ref{theory:scheme}) to treat the
$\left(\mathrm{H}_2\mathrm{O}\right)_n\mathrm{F}^-$ cluster. 
Beyond the complication of ligand-ligand H-bonding, 
we initially note that  the
hydration structures of $\mathrm{H_2O}- \mathrm{F}^-$ and
$\mathrm{H_2O}- \mathrm{Cl}^-$ are qualitatively different. In the
former case, classic OH-bond donation characterizes the clusters, while
in the latter case, more flexible but `\emph{dipole-dominated}'
configurations occur (Figure \ref{fig:snapshots}). Another interesting
observation is that, in contrast to
$\left(\mathrm{H}_2\mathrm{O}\right)_n\mathrm{Cl}^-$, the harmonic
approximation satisfactorily predicts free energy in the case of
$\left(\mathrm{H}_2\mathrm{O}\right)_n\mathrm{F}^-$ for $n \leq 4$
(right panel, Figure \ref{fig:cluster_formation_FE}).  

\begin{figure}[tbp]
\centering \boxed{\includegraphics[width=0.45\textwidth]{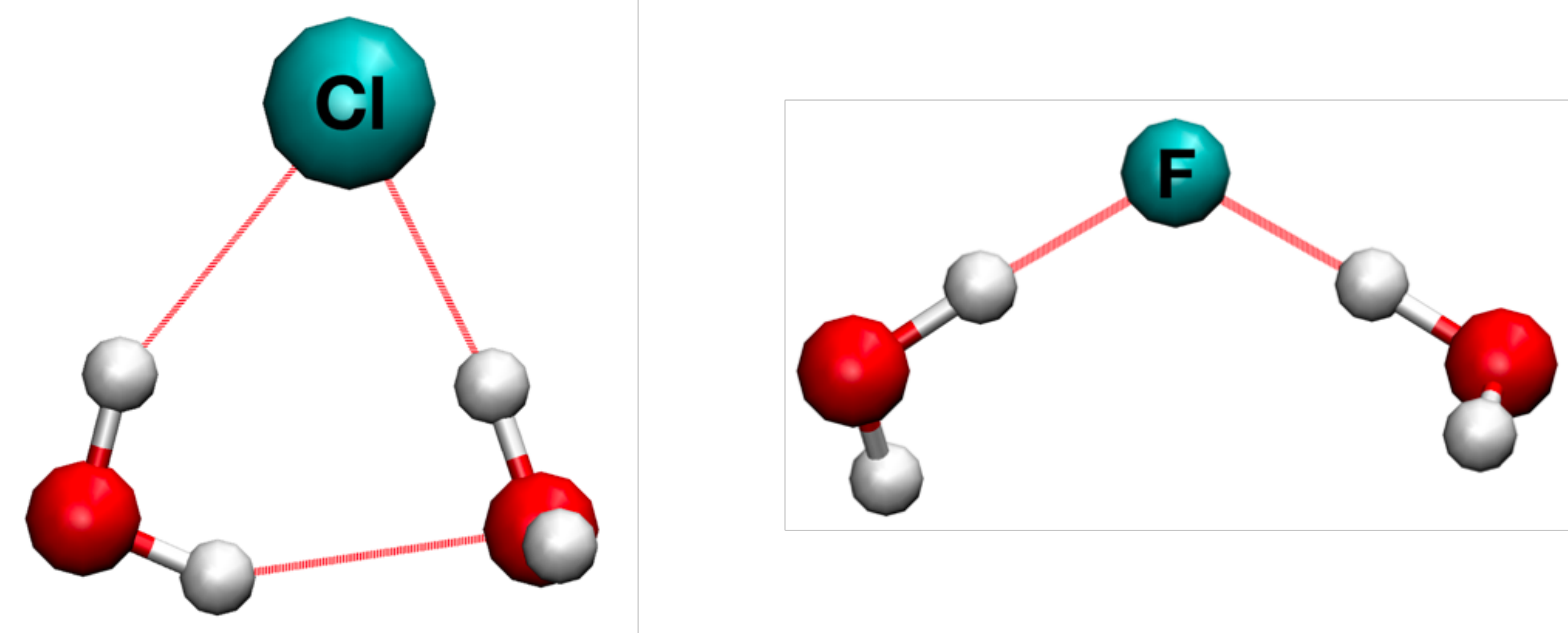}}
\caption{Optimized geometries for
$\left(\mathrm{H}_2\mathrm{O}\right)_2\mathrm{Cl}^-$ and
$\left(\mathrm{H}_2\mathrm{O}\right)_2\mathrm{F}^-$ obtained using
UPBE1PBE and B3LYP functionals respectively. The aug-cc-pvdz basis was
used in both cases. The
$\left(\mathrm{H}_2\mathrm{O}\right)_2\mathrm{Cl}^-$ cluster includes
inter-water H-bonding in contrast to
$\left(\mathrm{H}_2\mathrm{O}\right)_2\mathrm{F}^-$, consistent with (Ar
predissociation) spectroscopy studies for
halides.\cite{robertson2003molecular} A harmonic approximation
satisfactorily predicts the free energy for
$\left(\mathrm{H}_2\mathrm{O}\right)_2\mathrm{F}^-$ but not
$\left(\mathrm{H}_2\mathrm{O}\right)_2\mathrm{Cl}^-$.}    
\label{fig:opt_n2}
\end{figure}

These observations on the differences between $\mathrm{Cl}^-$  and
$\mathrm{F}^-$ can be explained as follows. Firstly, spectroscopic
studies\cite{robertson2003molecular} of halide-water clusters based on
Ar predissociation have identified vibrational bands that correspond to
inter-water H-bonding. The spectra revealed that those interactions
gradually weaken when going from
$\left(\mathrm{H}_2\mathrm{O}\right)_2\mathrm{I}^-$ to
$\left(\mathrm{H}_2\mathrm{O}\right)_2\mathrm{F}^-$ clusters. For
$\left(\mathrm{H}_2\mathrm{O}\right)_2\mathrm{F}^-$, that vibrational
band disappears, with water molecules separating entirely. This result
is consistent with our geometry optimization calculations for
$\left(\mathrm{H}_2\mathrm{O}\right)_2\mathrm{F}^-$ (Figure
\ref{fig:opt_n2}). The fact that a harmonic approximation reproduces the
experimental free energy at 298~K when applied to that minimum energy
structure indicates that inter-water H-bonding between waters is not
significant in $\left(\mathrm{H}_2\mathrm{O}\right)_n\mathrm{F}^-$. That
argument extends up to $n=4$ because no H-bond exists in those
energy-optimized structures. 

Secondly, previous \emph{ab initio} MD work \cite{tobias2001surface}
compared the computed IR absorption spectrum for
$\left(\mathrm{H}_2\mathrm{O}\right)_6\mathrm{Cl}^-$ with experimental
spectra obtained for
$\left(\mathrm{H}_2\mathrm{O}\right)_5\mathrm{Cl}^-$. That work
suggested that accounting for anharmonicity and coupling between modes
should be important for the treatment of inter-water H-bond dynamics in
aqueous clusters. Here, our step-wise scheme (Section
\ref{theory:scheme}) based on the ADMP approach naturally includes those
dynamical features. This quantitative treatment addresses the competing
inter-water H-bond dynamics in
$\left(\mathrm{H}_2\mathrm{O}\right)_n\mathrm{Cl}^-$ ($n\geq2$), leading
to cluster free energies in excellent agreement with experiments. The
earlier suggestion also explains why the harmonic approximation works
well for the case of $\left(\mathrm{H}_2\mathrm{O}\right)\mathrm{Cl}^-$,
where inter-water interactions are absent.            

\subsection{Free Energy of Hydration}

\begin{figure}[tbp]
\includegraphics[width=0.48\textwidth]{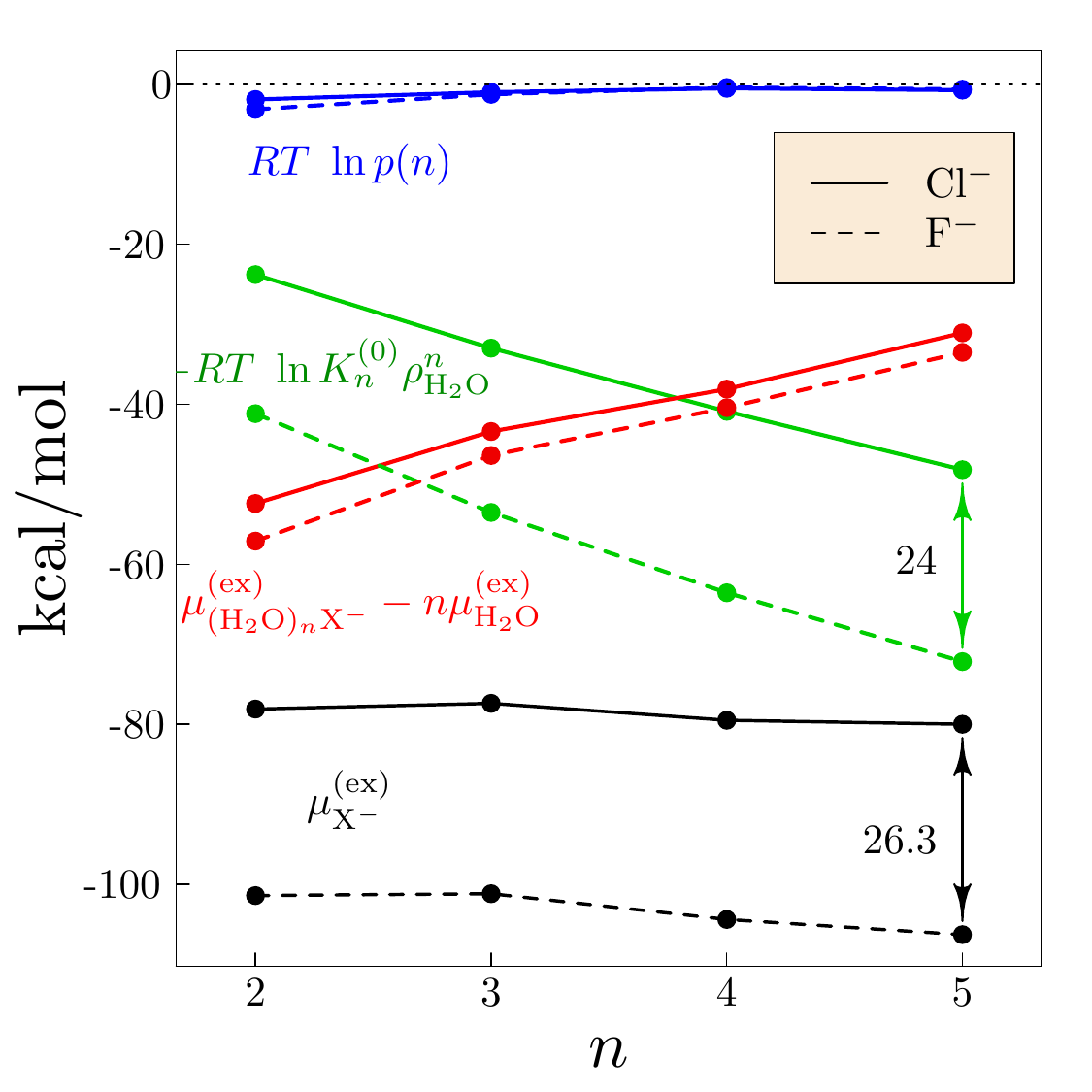}
\caption{Black: The excess free energy of hydration for X$^-$, evaluated
using Eq.~\eqref{eq:qct}. Red: The cluster hydration contribution to the
hydration free energy evaluated using the PCM model\cite{Tomasi:2005tc}
with configurations sampled from ADMP. Green: Isolated cluster free
energy from ADMP calculations (Figure~\ref{fig:cluster_formation_FE}).
Blue: The poly-dispersity contribution obtained from AIMD simulations.
Experimental values for the difference,
$\mu^{\mathrm{(ex)}}_{\mathrm{Cl}^{-}}$ -
$\mu^{\mathrm{(ex)}}_{\mathrm{F}^{-}}$, are $28$
kcal/mol\cite{friedman1973thermodynamics} and $29.8$
kcal/mol\cite{Marcus:1994ci} compared to $26.3$ kcal/mol here.
}
\label{fig:Cl_hydration_FE} 
\end{figure}

Summing the quasi-chemical components of Eq.\eqref{eq:qct}, we obtain an estimate of the net hydration free energy ($\mu^{\mathrm{(ex)}}_{\mathrm{X}^{-}}$, Figure~\ref{fig:Cl_hydration_FE}).
Note that the smallest contribution to the hydration free energy comes from $RT\ln p_{\mathrm{X}^{-}}(n)$, which
accounts for heterogeneity in occupancy of the inner solvation shell.
For the most probable occupation cases ($n=4,5$), this contribution is
inconsequential ($\approx -0.5~$kcal/mol) compared to the others. Next, the inner-shell interactions are built up by augmenting the isolated cluster free energy (evaluated at 1~atm, Figure~\ref{fig:cluster_formation_FE}) with $-nRT\ln 1354$, the ligand
replacement contribution that accounts for the actual density of ligands available in liquid phase at standard conditions, 1~g/cm$^3$. The full inner shell free energy contribution $\left( -RT\ln K^{(0)}_{n}\rho_{\mathrm{H_2O}}{}^{n}\right)$ increases in magnitude with $n$. The outer-shell contribution $\left(\mu^{\mathrm{(ex)}}_{\mathrm{(H_2O)_nX^-}}-n\mu^{\mathrm{(ex)
}}_{\mathrm{H_2O}}\right)$ is treated using the polarizable
continuum model (PCM) and decreases in magnitude with $n$. Finally, these contributions balance to produce a hydration free energy that is independent of $n$ (Figure \ref{fig:Cl_hydration_FE}). These evaluations agree reasonably well with experimental hydration free energies\cite{Marcus:1994ci} for Cl$^-$ ($-81.3$ kcal/mol) and F$^-$ ($-111.1$ kcal/mol). 

As an indication of the numerical sensitivity of 
such calculations, 
we note a previous QCT treatment that arrived at a value  
of {$-228$~kcal/mol} for the LiF (aq) 
pair.\cite{muralidharan2018quasi}  For comparison, 
an AIMD calculation based on an electron 
density functional obtained $-240$~kcal/mol.\cite{Duignan:2017iha}  These 
two results 
bracket experimental values, nearly equally.  When the 
latter result was corrected \emph{ex post facto} by MP2
calculations on a finite solution fragment 
for the F$^-$ case, the corrected value splits 
the indicated difference.  This comparison 
suggests the importance of detailed treatment of 
dispersion effects for anions, though not necessarily for cations.\cite{Soniat:2015}

Since QCT explicitly discriminates between inner- and outer-shell 
effects, comparison of the different free energy contributions to 
hydration of
Cl$^-$ and F$^-$ leads to some interesting observations
(Figure~\ref{fig:Cl_hydration_FE}). The cluster hydration contributions,
$\left(\mu^{\mathrm{(ex)}}_{\mathrm{(H_2O)_nX^-}}-n\mu^{\mathrm{(ex)
}}_{\mathrm{H_2O}}\right)$, are nearly identical. This quantitatively
supports the intuition that long-ranged interactions, predominantly
electrostatic, do not discriminate between Cl$^-$ and F$^-$. In
contrast, the inner-shell interactions, $-RT\ln
K^{(0)}_{n}\rho_{\mathrm{H_2O}}{}^{n}$, differ by 24~kcal/mol,
essentially making up the difference (26.3~kcal/mol) in the hydration
free energies. Clearly, disparities in the size and electronic structure
of these ions contribute to the differences reported here, but those
effects are localized to the inner solvation shell. Thus, the treatment
of specific ion effects can focus primarily on accurate evaluation of
inner-shell contributions to free energy through rigorous treatment of
dynamics in small clusters, as demonstrated here.    

\section{Discussion and Conclusions}
\label{sec:conclusions}
These results confirm again that a foremost difficulty in treating
hydrated anions such as Cl$^-$(aq) is significant water-water H-bonding
within the inner solvation shell (Figure~\ref{fig:opt_n2}). 
Nevertheless, the hydration structures of \emph{one-water molecule}
cases, $\mathrm{H_2O}- \mathrm{F}^-$ and $\mathrm{H_2O}- \mathrm{Cl}^-$,
differ qualitatively (Figures~\ref{fig:snapshots} and \ref{fig:angle}),
with the latter complex being qualitatively a \emph{dipole-dominated}
structure. Of course, this feature facilitates inter-ligand H-bonding
for the Cl$^-$(aq) case. The ADMP computational tool works
satisfactorily for dynamics of the isolated inner-shell clusters that
are treated in QCT. But  details of the trajectory thermostat
algorithm  deserve further elucidation (Supporting Information).  A
one-ligand stepwise evaluation of the cluster free energies works
surprisingly well, and those computational results agree nicely with
experimental cluster free energies
(Figure~\ref{fig:cluster_formation_FE}).

Collection of the several factors contributing to the QCT free energy
(Figure \ref{fig:Cl_hydration_FE}) produces net results in good
agreement with experimental
tabulations.\cite{friedman1973thermodynamics,Marcus:1994ci} More
decisively, the difference $\mu^{\mathrm{(ex)}}_{\mathrm{Cl}^{-}}$ -
$\mu^{\mathrm{(ex)}}_{\mathrm{F}^{-}}$ is not subject to an ambiguity
of a \emph{potential of the phase} (or `surface potential'). 
Thermodynamic results then provide an unambiguous test for the
difference. Here, the QCT result for the difference is about 2-3~kcal/mol
smaller than the experimental value for that difference, about
-29~kcal/mol. This discrepancy is likely due to use of the PCM treatment for the
cluster-hydration contribution, which treats long-range interactions.
Thus, differential outer-shell
structuring not considered by PCM is likely a principal contribution 
to the difference.  The AIMD results presented here
(Figure~\ref{fig:Neighborship_H}) characterize that outer-shell
structure.  Note that the QCT approach uses the PCM approximation  in a context
where it is insensitive to adjustment of dielectric radii, so the
discrepancy identified here is likely to be intrinsic to the PCM model.

The QCT analysis assigns (Figure \ref{fig:Cl_hydration_FE}) 
90\% of the hydration free energy difference $\mu^{\mathrm{(ex)}}_{\mathrm{Cl}^{-}}$
- $\mu^{\mathrm{(ex)}}_{\mathrm{F}^{-}}$  to inner-shell contributions
deriving from the isolated hydrous cluster. This result is 
an important
idea to keep in mind in addressing Hofmeister effects and 
selectivity of
ion channels.\cite{Merchant:2009fq,Willow:2017ij}  Precision in
analyzing single-ion free energies should assist in understanding
selectivity in ion channels by establishing end-point free energies
bracketing a path for ion transit.

\section{Methods}\label{sec:methods}\label{methods}
\subsection{ADMP}
We evaluated molecular dynamics of the isolated 
(H$_2$O)$_n\mathrm{Cl}^-$ clusters for $1 \leq n \leq 5$ using atom-centered
basis sets and the density matrix propagation (ADMP)
approach\cite{schlegel2002ab} available through the Gaussian09 software
package.\cite{g09} The UPBE1PBE density functional was utilized with the
aug-cc-pvdz basis set. The fictitious masses that couple electron motion
with the nuclei was set to 0.1~amu. The initial velocities of the
individual atoms were sampled randomly from a Boltzmann distribution,
and the initial density matrix velocity was chosen to be zero. With a
time step of 0.1~fs, dynamic simulations were carried out for 1~ps at
300 K for all cluster sizes, and the last 0.5~ps were used for analysis.

 One reservation in using Gaussian09 for dynamics is that
description of the only implemented thermostat option has been
sketchy.\cite{schlegel2002ab} Operationally, this algorithm constrains
the total nuclear kinetic energy of the system. 
Intuitive, non-canonical
thermostats are known to cause problems in  some cases, as in 
the flying
ice cube effect.\cite{harvey1998flying}  Nevertheless, due to lack of
alternatives in Gaussian09, we used that thermostat to maintain the
system at 300~K. 

Though reservations with that thermostat should be borne
in mind, they do not appear to be serious here. First, the atomic
velocities are verified to be Maxwell-Boltzmann distributed 
(SI, Figure
S2). In addition, we performed 10~ps of \emph {ab initio} 
molecular dynamics at
300~K with the Nose-Hoover thermostat\cite{Nose} for
$\left(\mathrm{H}_2\mathrm{O}\right)_2\mathrm{Cl}^-$ using
CP2K.\cite{cp2k2005quickstep} The PBE density
functional\cite{perdew1996generalized} was utilized with
GTH\cite{goedecker1996separable} pseudopotentials
in the Gaussian and plane wave
schemes.\cite{lippert1999gaussian} Molecularly optimized  
DZVP-MOLOPT-GTH\cite{vandevondele2007gaussian} basis sets were obtained
from the CP2K website.   The structures sampled by this CP2K
trajectory were then analyzed with Gaussian09 using the 
UPBE1PBE density
functional and the aug-cc-pvdz basis set. The isolated cluster free
energy obtained this way is within 0.1 kcal/mol of the ADMP trajectory
result. A detailed comparison of the cluster simulation methods
discussed above is provided with the Supporting Information (SI, Figure S3).

\subsection{PCM for ADMP sampled clusters}
The outer-shell cluster contribution $\left(
\mu^{\mathrm{(ex)}}_{\mathrm{(H_2O)}_n\mathrm{Cl}^-}-n\mu^{\mathrm{(ex)
}}_{\mathrm{H_2O}}
\right)$ to the hydration free energy is treated using the polarizable
continuum model\cite{Tomasi:2005tc} (PCM) in Gaussian09.\cite{g09}
Configurations that obey the clustering constraint were extracted from
the last 1~ps of the ADMP trajectory. The sampled 
geometrical structures
were subjected to two (2) single point electronic calculations,
separately, one for the isolated cluster and a second with the external 
(dielectric)
medium described by the PCM tool. The difference, $\bar \varepsilon_j$,  is
employed in computing
\begin{eqnarray}
\mu^{{\mathrm{(ex)}}}_{\mathrm{(H_2O)}_n\mathrm{Cl}^-} = -RT
\ln \left\lbrack
\left(\frac{1}{N}\right)\sum\limits_{j=1}^{N}\me^{-{\bar\varepsilon}_j/RT} 
\right\rbrack ,
\label{eq:widom}
\end{eqnarray}
where $N$ is the number of configurations from the simulation stream
that obey the clustering constraint. Eq.~\eqref{eq:widom} 
corresponds to
the potential distribution theorem (PDT)
approach,\cite{beck2006potential,asthagiri2010ion} recognizing that
thermal fluctuations implicit in the PCM\cite{Tomasi:2005tc} approach
complete the PDT averaging.

\subsection{AIMD}
A system consisting of a single halide ion (Cl$^-$ or F$^-$) and 64
waters was simulated using the VASP AIMD simulation
package.\cite{kresse1993ab,kresse1996efficient}  A 
wide range of alternative electron density functionals
are available, in principal,\cite{Gillan:2016ff} for such
calculations.  Here we chose the PW91
generalized gradient approximation principally for consistency with
previous calculations.\cite{Chaudhari:2017gs} The plane wave basis had a
high kinetic energy cutoff of 400~eV.

A cubic cell of 1.24~nm was used to set a satisfactory density of the
water. The calculation adopted the  Nos\'e thermostat
procedure\cite{Nose} at 350~K.  With respect to  the ADMP calculations
with $T$ = 300~K and with the water density set, that 
temperature adjustment
amounts to a scaling of the potential energy by 300/350.  That scaling
at constant density avoids glassy behavior that can result
otherwise.\cite{Rempe:water}

A time step of 0.5~fs was chosen to realize a 100~ps trajectory. The
last 50~ps were used for analysis. The radial distribution functions,
g$_{\mathrm{H}|\mathrm{X}} (r)$, and the inner-shell occupancy
probabilities, $p_{\mathrm{X}^{-}}(n)$, were evaluated on that basis.
The radial distribution functions 
(Figure \ref{fig:Neighborship_H}) provide 
qualitative guidance to the present statistical theory, but not 
numerical input.  The occupancy
probabilities, $p_{\mathrm{X}^{-}}(n)$, contribute 
to Eq.~\eqref{eq:qct} but at the boundary of significance for these 
studies (Figure~\ref{fig:Cl_hydration_FE}).

\section{Acknowledgements}
This work was performed, in part, at the Center for
Integrated Nanotechnologies, an Office of Science User 
Facility operated
for the U.S. Department of Energy (DOE) Office of Science. The 
work was
supported by Sandia National Laboratories (SNL) LDRD program. SNL
is a multi-mission laboratory managed and operated by National
Technology and Engineering Solutions of Sandia, LLC., a wholly owned
subsidiary of Honeywell International, Inc., for the U.S. 
DOE's National
Nuclear Security Administration under contract DE-NA-0003525. 
The views
expressed in the article do not necessarily represent the views of the
U.S. DOE or the United States Government.  LRP thanks Dilip Asthagiri
and Diego Gomez for helpful discussions.

\newpage

\section*{Supporting Information}

\bigskip
\subsection*{Geometry optimizations}

\bigskip
\begin{figure}[!htbp]
\includegraphics[width=0.44\textwidth]{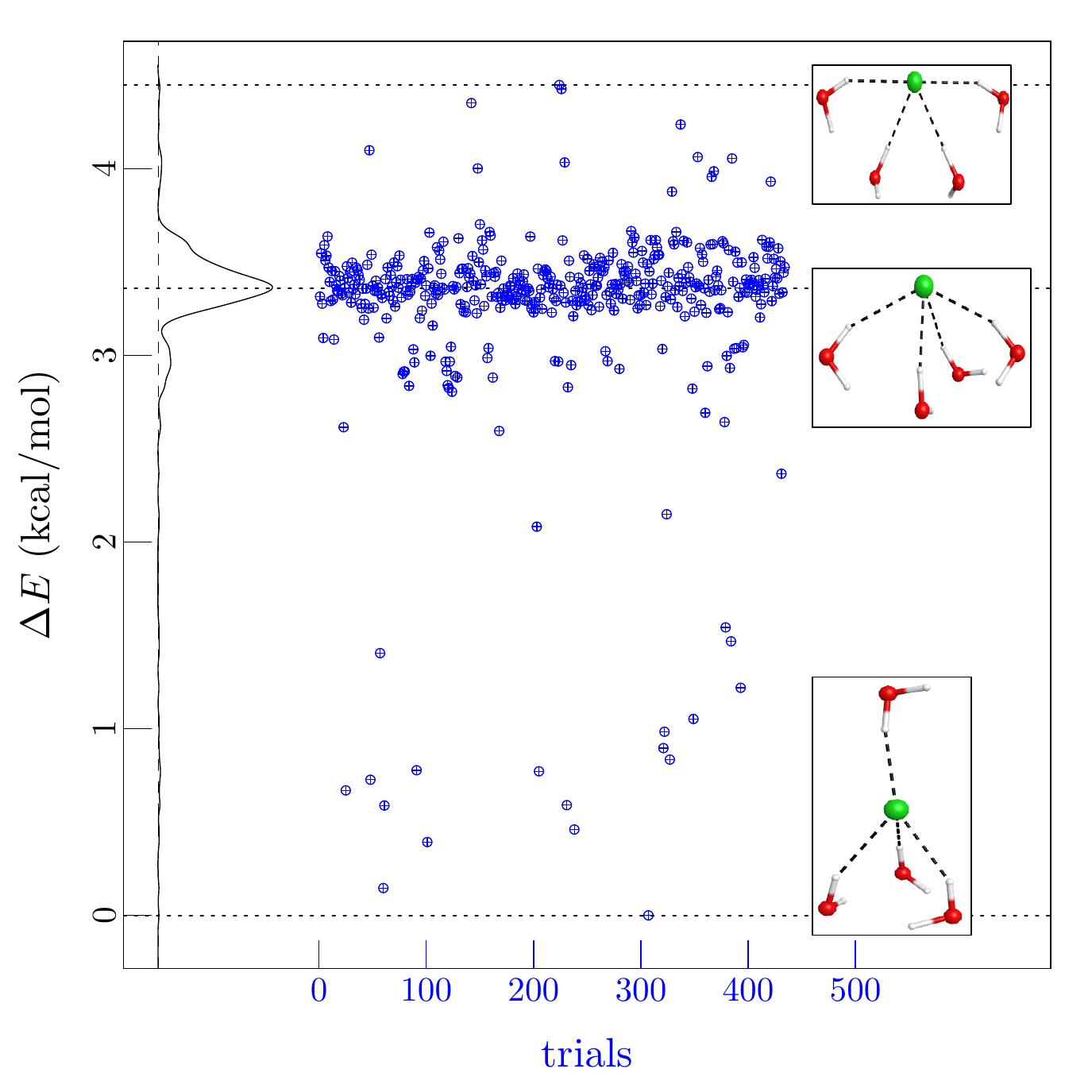}
\hfill
\includegraphics[width=0.44\textwidth]{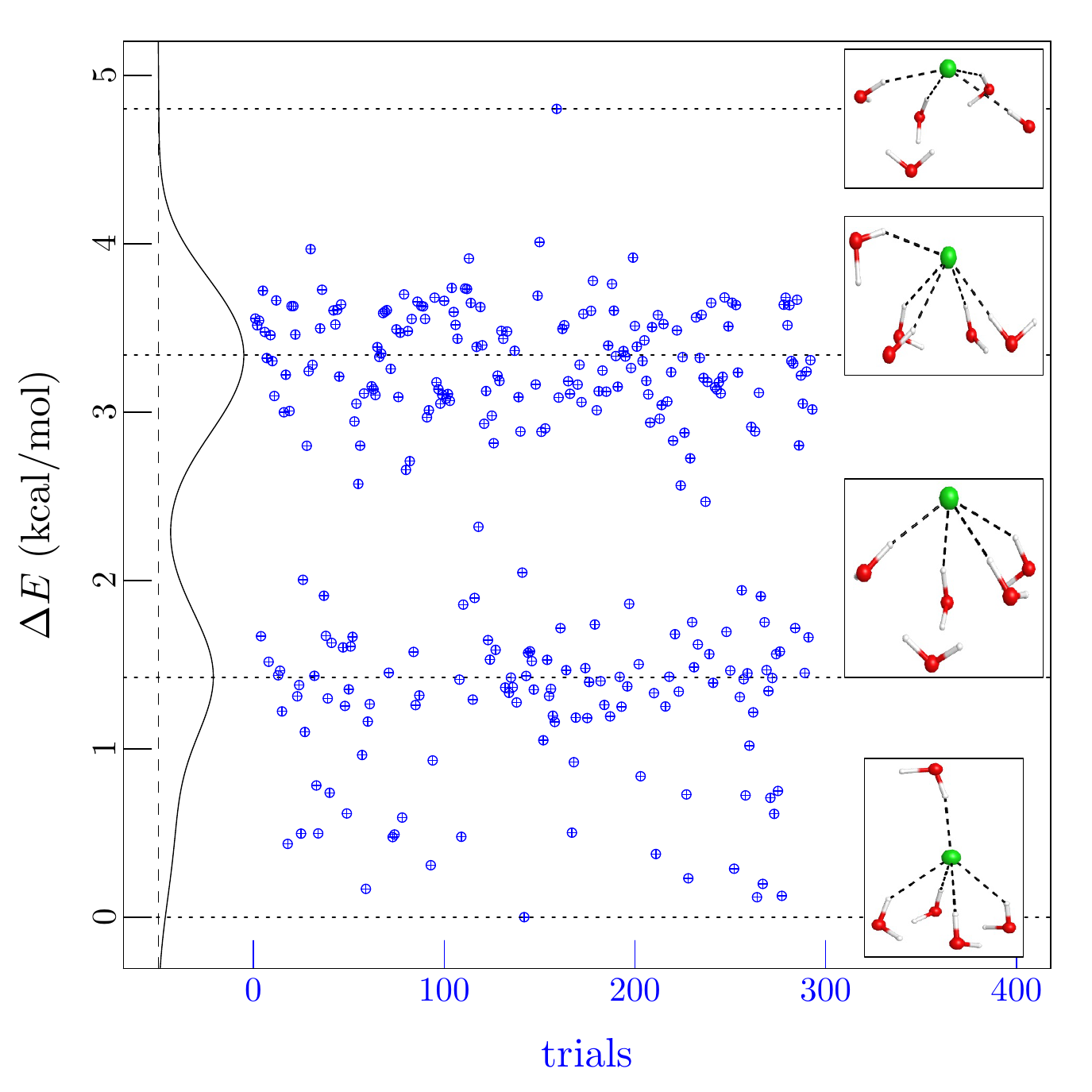}
\label{fig:Cl_sample_N5_2}
\caption{Blue dots: Electronic energy of optimized $\left(\mathrm{H}_2\mathrm{O}\right)_4\mathrm{Cl}^-$ (left) and $\left(\mathrm{H}_2\mathrm{O}\right)_5\mathrm{Cl}^-$ (right) clusters with 
initial configurations sampled from bulk phase AIMD. Black curve on the left shows 
the
distribution of those energies. The lowest optimum energy  (bottom inset) is about 
4-5 kcal/mol lower in energy than the highest optimum energy (top inset). The harmonic approximation for estimation of cluster free energy is based on the lowest energy optimum. That structure is also used as a starting configuration for molecular dynamics with ADMP. }
\end{figure}
\FloatBarrier
\newpage
\subsection*{Thermostat testing}
\begin{figure}[!htbp]
\begin{center} \includegraphics[width=0.4\textwidth]{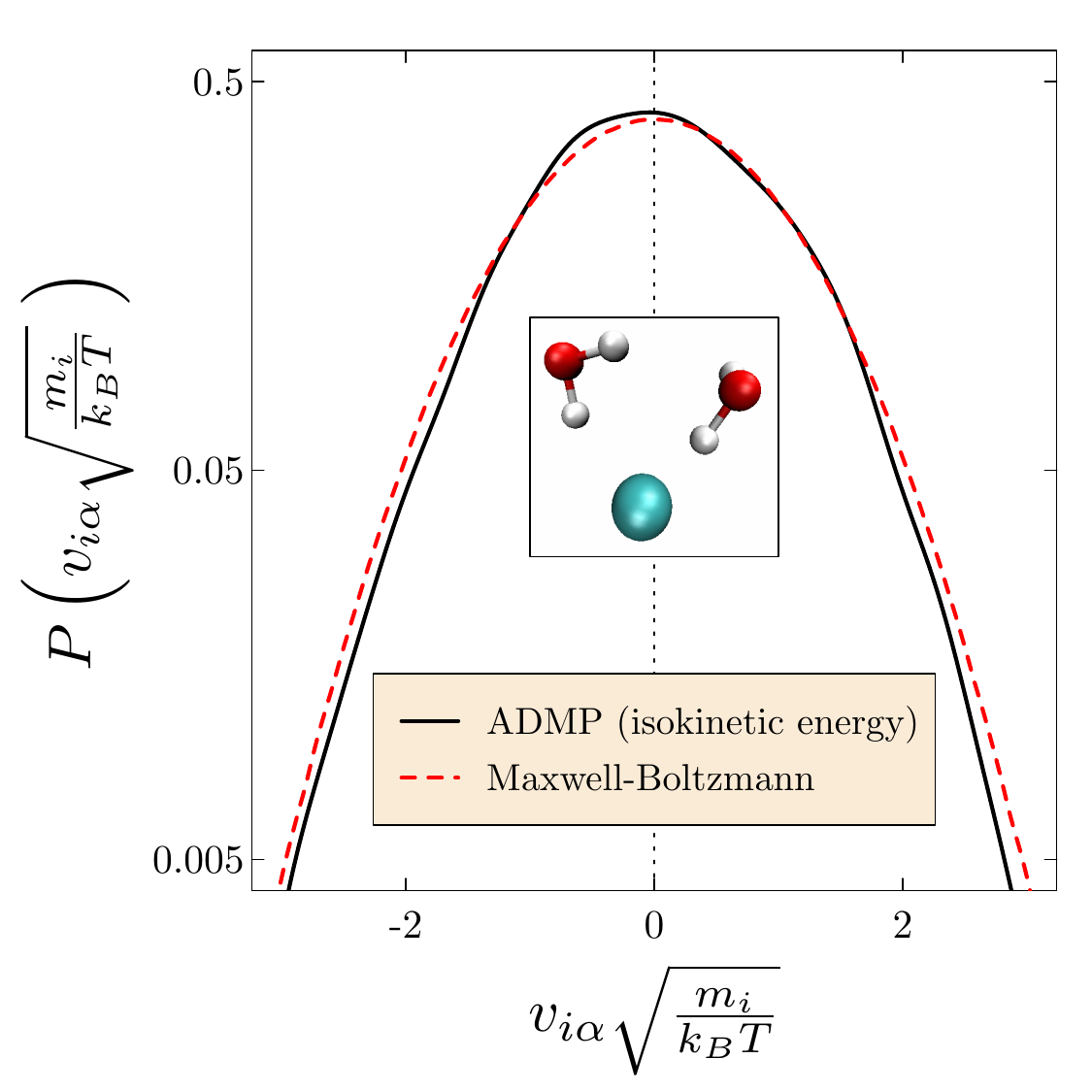} \end{center}
\caption{Testing the thermostat implemented with the ADMP approach\cite{schlegel2002ab} in the
Gaussian09 software package\cite{g09}
for the cluster $\left(\mathrm{H}_2\mathrm{O}\right)_2\mathrm{Cl}^-$ at
$T$=300~K. The UPBE1PBE density functional was utilized with the
aug-cc-pvdz basis set. Variable $\alpha = x, y, z$  indexes the Cartesian
components of atomic velocity. The observed distribution of atomic
velocities reasonably matches the Maxwell-Boltzmann distribution.} 
\label{fig:Admp_E} 
\end{figure}

\newpage
\subsection*{Comparison of cluster simulation methods}
\subsubsection*{Gaussian09 trajectory:} The ADMP approach 
utilized the UPBE1PBE 
density functional with the aug-cc-pvdz basis set. The 
reservation in using Gaussian09 for dynamics is that the only 
available thermostat option is not fully described. 
Nevertheless, the velocity distribution 
obtained here (Figure \ref{fig:Admp_E}) matches 
Maxwell-Boltzmann distribution. For further testing, 
we use CP2K, as described below.\bigskip

\subsubsection*{CP2K trajectory:} CP2K allows molecular dynamics 
with the Nose-Hoover thermostat.\cite{Nose} The PBE density 
functional\cite{perdew1996generalized} was utilized with 
pseudopotentials proposed by Goedecker, Teter and Hutter 
(GTH\cite{goedecker1996separable}) in the hybrid Gaussian and plane 
wave scheme.\cite{lippert1999gaussian} Molecularly optimized basis 
sets denoted as DZVP-MOLOPT-GTH\cite{vandevondele2007gaussian} were 
obtained from the CP2K website. The trajectory obtained is then 
analysed in Gaussian09 using the UPBE1PBE density functional with 
the aug-cc-pvdz basis set.

\begin{figure}[t]
\includegraphics[width=0.4\textwidth]{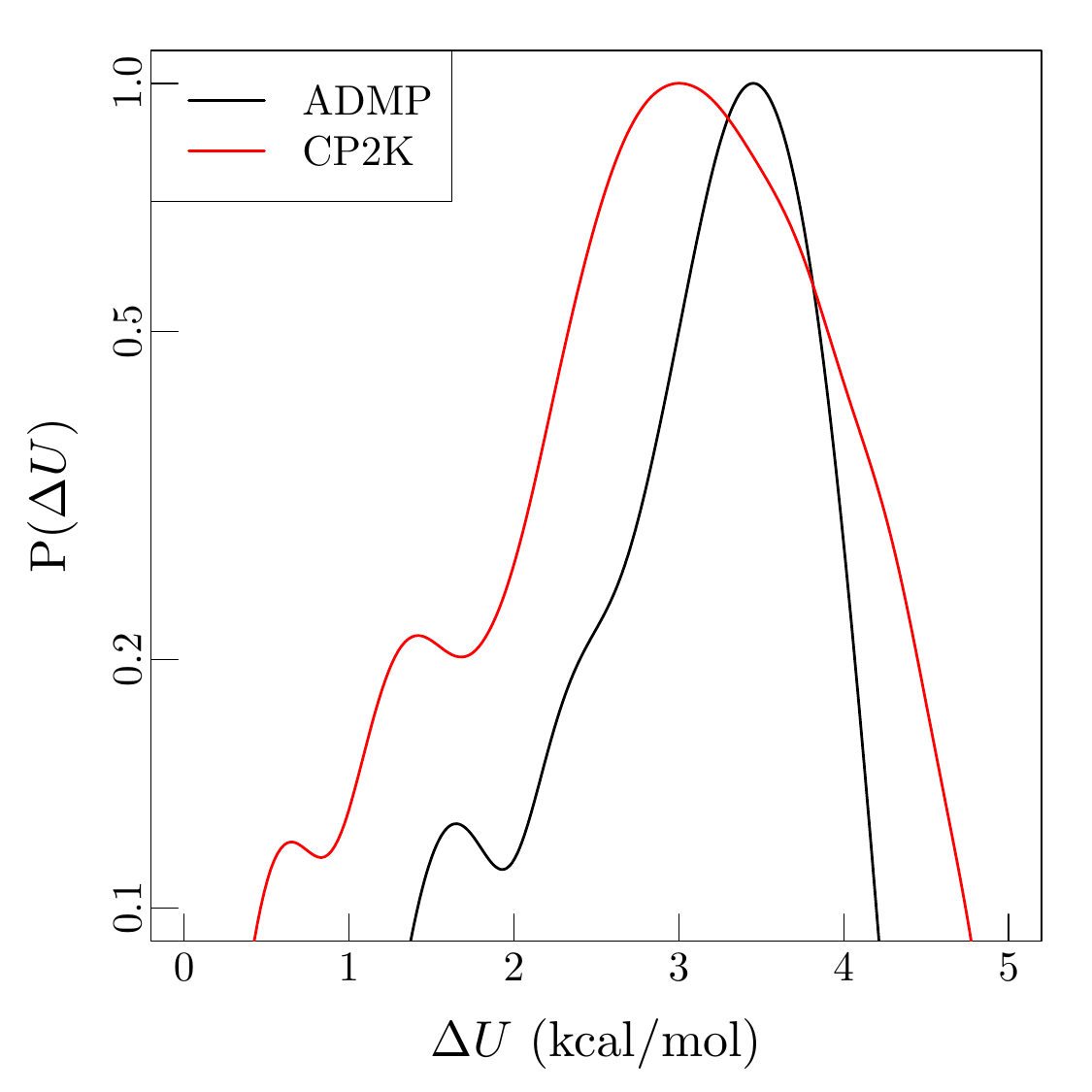}
\includegraphics[width=0.4\textwidth]{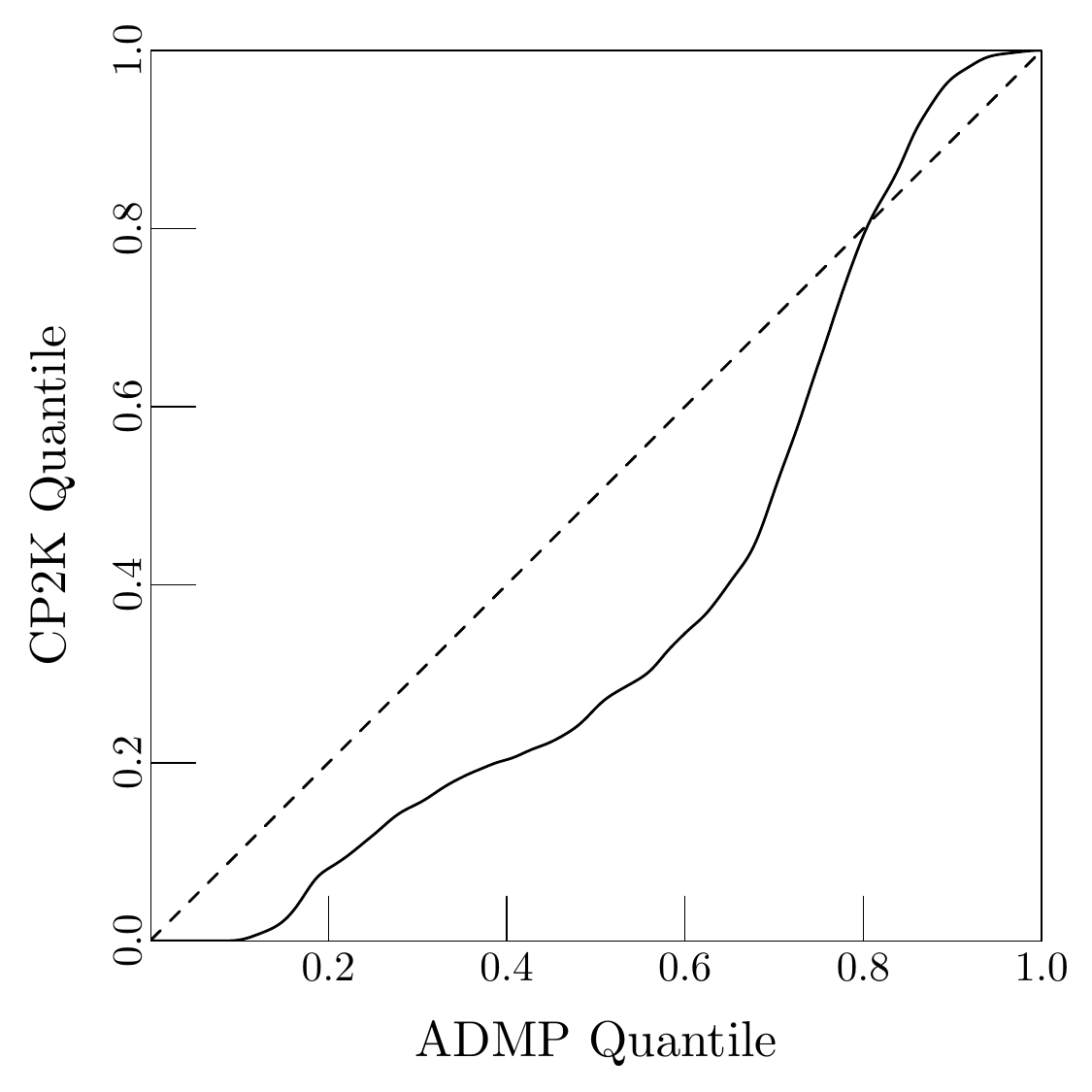}
\label{Q-Q}
\caption{Upper: Comparison of $\Delta U$ samples for 
$\left(\mathrm{H}_2\mathrm{O}\right)_2\mathrm{Cl}^-$ obtained using 
the ADMP approach\cite{schlegel2002ab} (black) in the 
Gaussian09 software package 
with the mixed Gaussian and plane wave approach (red) in the CP2K 
package.\cite{cp2k2005quickstep} $\Delta U = E\left(\gamma_n\sigma 
\right) - E\left(\gamma_{n-1}\sigma \right) - E\left(\gamma\sigma 
\right) + E(\sigma)~$, defined in the main text.
Lower: The standard Q-Q plot for the distribution shown in the left 
panel.}
\end{figure}
\vfill
\clearpage
\bibliography{Bibliography}

\begin{thebibliography}{100}
\expandafter\ifx\csname url\endcsname\relax
  \def\url#1{\texttt{#1}}\fi
\expandafter\ifx\csname urlprefix\endcsname\relax\def\urlprefix{URL }\fi
\expandafter\ifx\csname href\endcsname\relax
  \def\href#1#2{#2} \def\path#1{#1}\fi

\bibitem{Jungwirth:2006gu}
P.~Jungwirth, D.~J. Tobias, {Specific Ion Effects at the Air/Water Interface},
  Chem. Rev. 106 (2006) 1259--1281.

\bibitem{Kalcher:2009js}
I.~Kalcher, D.~Horinek, R.~R. Netz, J.~Dzubiella, {Ion specific correlations in
  bulk and at biointerfaces}, J. Phys. Cond. Mat. 21 (2009) 424108.

\bibitem{Zhang:2010gr}
Y.~Zhang, P.~S. Cremer, {Chemistry of Hofmeister Anions and Osmolytes}, Annu
  Rev Phys Chem 61 (2010) 63--83.

\bibitem{Kunz:2010fy}
W.~Kunz, {Specific ion effects in colloidal and biological systems}, Curr. Op.
  Coll. {\&} Interf. Sci. 15 (2010) 34--39.

\bibitem{dpars11}
D.~F. Parsons, M.~Bostr{\"o}m, P.~L. Nostro, B.~W. Ninham, {Hofmeister effects:
  interplay of hydration, nonelectrostatic potentials, and ion size}, Phys.
  Chem. Chem. Phys. 13 (2011) 12352.

\bibitem{Pollard:2016ei}
T.~P. Pollard, T.~L. Beck, {Toward a quantitative theory of Hofmeister
  phenomena: From quantum effects to thermodynamics}, Curr. Opin. Coll. {\&}
  Interf. Sci. 23 (2016) 110--118.

\bibitem{SPERELAKIS2012121}
N.~Sperelakis, {Origin of Resting Membrane Potentials}, in: Cell Physiology
  Source Book (Fourth Edition), Academic Press, 2012, pp. 121 -- 145.

\bibitem{you2014comparison}
X.~You, M.~I. Chaudhari, L.~R. Pratt, {Comparison of mechanical and
  thermodynamical evaluations of electrostatic potential differences between
  electrolyte solutions}, in: P.~Lo~Nostro, B.~W. Ninham (Eds.), Aqua
  Incognita: Why ice floats on water and Gallileo 400 years on, Connor Court
  Publishing Pty Ltd, 2014, pp. 434--442.

\bibitem{pratt1992contact}
L.~R. Pratt, Contact potentials of solution interfaces: phase equilibrium and
  interfacial electric fields, J. Phys. Chem. 96 (1992) 25--33.

\bibitem{Leung:2009dx}
K.~Leung, S.~B. Rempe, O.~A. von Lilienfeld, {Ab initio molecular dynamics
  calculations of ion hydration free energies}, J. Chem. Phys. 130~(20) (2009)
  204507--204517.

\bibitem{lyklema2017interfacial}
J.~Lyklema, {Interfacial Potentials: Measuring the Immeasurable?}, Substantia
  1~(2).

\bibitem{doyle2019importance}
C.~C. Doyle, Y.~Shi, T.~L. Beck, The importance of the water molecular
  quadrupole for estimating interfacial potential shifts acting on ions near
  the liquid-vapor interface, The Journal of Physical Chemistry B.

\bibitem{Jungwirth:2002et}
P.~Jungwirth, D.~J. Tobias, {Chloride Anion on Aqueous Clusters, at the
  Air$-$Water Interface, and in Liquid Water:~ Solvent Effects on Cl$^-$
  Polarizability}, J. Phys. Chem. A 106 (2002) 379--383.

\bibitem{Jungwirth:2002eu}
P.~Jungwirth, D.~J. Tobias, {Ions at the Air/Water Interface}, J. Phys. Chem. B
  106 (2002) 6361--6373.

\bibitem{Tongraar:2003bd}
A.~Tongraar, B.~Michael~Rode, {The hydration structures of {F{$^-$}and
  Cl{$^-$}} investigated by {\emph{ab initio}} {QM/MM} molecular dynamics
  simulations}, Phys. Chem. Chem. Phys. 5 (2003) 357--362.

\bibitem{Heuft:2005jt}
J.~M. Heuft, E.~J. Meijer, {Density functional theory based molecular-dynamics
  study of aqueous iodide solvation}, J. Chem. Phys. 123 (2005) 094506--6.

\bibitem{Heuft:2005kxa}
J.~M. Heuft, E.~J. Meijer, {Density functional theory based molecular-dynamics
  study of aqueous fluoride solvation}, J. Chem. Phys. 122 (2005) 094501--8.

\bibitem{Heuft:2003iva}
J.~M. Heuft, E.~J. Meijer, {Density functional theory based molecular-dynamics
  study of aqueous chloride solvation}, J. Chem. Phys. 119 (2003) 11788--11791.

\bibitem{Ho:2009bga}
M.-H. Ho, M.~L. Klein, I.~F.~W. Kuo, {Bulk and Interfacial Aqueous Fluoride: An
  Investigation via First Principles Molecular Dynamics}, J. Phys. Chem. A 113
  (2009) 2070--2074.

\bibitem{Raugei:2002gf}
S.~Raugei, M.~L. Klein, {An ab initio study of water molecules in the bromide
  ion solvation shell}, J. Chem. Phys. 116 (2002) 196--8.

\bibitem{cabarcos1999microscopic}
O.~M. Cabarcos, C.~J. Weinheimer, J.~M. Lisy, S.~S. Xantheas, Microscopic
  hydration of the fluoride anion, J. Chem. Phys. 110 (1999) 5--8.

\bibitem{bankura2013hydration}
A.~Bankura, V.~Carnevale, M.~L. Klein, Hydration structure of salt solutions
  from ab initio molecular dynamics, J. Chem. Phys. 138 (2013) 014501.

\bibitem{Perera:1994ii}
L.~Perera, M.~L. Berkowitz, {Structures of Cl$^-$(H$_2$O)$_n$ and
  F$^-$(H$_2$O)$_n$ (n=2,3,...,15) clusters. Molecular dynamics computer
  simulations}, J. Chem. Phys. 100 (1994) 3085--3093.

\bibitem{Sremaniak:1994wm}
L.~S. Sremaniak, L.~Perera, M.~L. Berkowitz, {Enthalpies of formation and
  stabilization energies of Br$^-$(H$_2$O)$_n$ ($n$=1,2,\ldots , 15) clusters.
  Comparisons between molecular dynamics computer simulations and experiments},
  Chem. Phys. Lett. 218 (1994) 377--382.

\bibitem{Herce:2005dp}
D.~H. Herce, L.~Perera, T.~A. Darden, C.~Sagui, {Surface solvation for an ion
  in a water cluster}, J. Chem. Phys. 122 (2005) 024513--11.

\bibitem{Eggimann:2007cp}
B.~L. Eggimann, J.~I. Siepmann, {Size Effects on the Solvation of Anions at the
  Aqueous Liquid-Vapor Interface}, J. Phys. Chem. C 112 (2007) 210--218.

\bibitem{Bostrom:2005km}
M.~Bostr{\"o}m, W.~Kunz, B.~W. Ninham, {Hofmeister effects in surface tension
  of aqueous electrolyte solution}, Langmuir 21~(6) (2005) 2619--2623.

\bibitem{asthagiri2010ion}
D.~Asthagiri, P.~Dixit, S.~Merchant, M.~Paulaitis, L.~Pratt, S.~B. Rempe,
  S.~Varma, Ion selectivity from local configurations of ligands in solutions
  and ion channels, Chem. Phys. Lett. 485 (2010) 1--7.

\bibitem{rogers2012:structural}
D.~M. Rogers, D.~Jiao, L.~R. Pratt, S.~B. Rempe, Structural models and
  molecular thermodynamics of hydration of ions and small molecules, in: Annu
  Rep Comput Chem, Vol.~8, Elsevier, 2012, pp. 71--127.

\bibitem{Rogers:2010gh}
D.~M. Rogers, T.~L. Beck, {Quasichemical and structural analysis of polarizable
  anion hydration}, J. Chem. Phys. 132 (2010) 014505--13.

\bibitem{pratt1999quasi}
L.~R. Pratt, S.~B. Rempe, Quasi-chemical theory and implicit solvent models for
  simulations, in: Simulation and Theory of Electrostatic Interactions in
  Solution. Computational Chemistry, Biophysics, and Aqueous Solutions, Vol.
  492 of AIP Conference Proceedings, American Institute of Physics, Vol. 492,
  AIP, 1999, pp. 172--201.

\bibitem{rempe2000hydration}
S.~B. Rempe, L.~R. Pratt, G.~Hummer, J.~D. Kress, R.~L. Martin, A.~Redondo,
  {The Hydration Number of Li{$^+$} in Liquid Water}, J. Am. Chem. Soc. 122
  (2000) 966--967.

\bibitem{paliwal2006analysis}
A.~Paliwal, D.~Asthagiri, L.~R. Pratt, H.~S. Ashbaugh, M.~E. Paulaitis, An
  analysis of molecular packing and chemical association in liquid water using
  quasichemical theory, J. Chem. Phys. 124 (2006) 224502.

\bibitem{beck2006potential}
T.~L. Beck, M.~E. Paulaitis, L.~R. Pratt, The Potential Distribution Theorem
  and Models of Molecular Solutions, Cambridge University Press, 2006.

\bibitem{shah2007balancing}
J.~Shah, D.~Asthagiri, L.~Pratt, M.~Paulaitis, Balancing local order and
  long-ranged interactions in the molecular theory of liquid water, J. Chem.
  Phys. 127 (2007) 144508.

\bibitem{Anonymous:2009dc}
S.~Chempath, L.~R. Pratt, {Distribution of Binding Energies of a Water Molecule
  in the Water Liquid$-$Vapor Interface}, J. Phys. Chem. B 113 (2009)
  4147--4151.

\bibitem{muralidharan2018quasi}
A.~Muralidharan, L.~R. Pratt, M.~I. Chaudhari, S.~B. Rempe, {Quasi-Chemical
  Theory With Cluster Sampling From Ab Initio Molecular Dynamics: Fluoride
  (F$^-$) Anion Hydration}, J. Phys. Chem. A 122 (2018) 9806--9812.

\bibitem{pratt2003self}
L.~R. Pratt, H.~S. Ashbaugh, Self-consistent molecular field theory for packing
  in classical liquids, Phys. Rev. E 68 (2003) 021505.

\bibitem{Weber:2011hd}
V.~Weber, S.~Merchant, D.~Asthagiri, {Communication: Regularizing binding
  energy distributions and thermodynamics of hydration: Theory and application
  to water modeled with classical and ab initio simulations}, J. Chem. Phys.
  135 (2011) 181101.

\bibitem{sabo2006:H2}
D.~Sabo, S.~Rempe, J.~Greathouse, M.~Martin, Molecular studies of the
  structural properties of hydrogen gas in bulk water, Mol. Simulat. 32~(3-4)
  (2006) 269--278.

\bibitem{sabo2008hydrogen}
D.~Sabo, S.~Varma, M.~G. Martin, S.~B. Rempe, Studies of the thermodynamic
  properties of hydrogen gas in bulk water, J. Phys. Chem. B 112~(3) (2008)
  867--876.

\bibitem{clawson2010:H2}
J.~S. Clawson, R.~T. Cygan, T.~M. Alam, K.~Leung, S.~B. Rempe, Ab initio study
  of hydrogen storage in water clathrates, J. Comput. Theor. Nanosci. 7~(12)
  (2010) 2602--2606.

\bibitem{jiao2012combined}
D.~Jiao, S.~B. Rempe, Combined density functional theory {(DFT)} and continuum
  calculations of {p$K_a$} in carbonic anhydrase, Biochem. 51~(30) (2012)
  5979--5989.

\bibitem{jiao2011co2}
D.~Jiao, S.~B. Rempe, {CO{$_2$} Solvation Free Energy Using Quasi-Chemical
  Theory}, J. Chem. Phys. 134 (2011) 224506.

\bibitem{Chaudhari:2015jq}
M.~I. Chaudhari, D.~Sabo, L.~R. Pratt, S.~B. Rempe, {Hydration of Kr(aq) in
  Dilute and Concentrated Solutions}, J. Phys. Chem. B 119 (2015) 9098--9102.

\bibitem{varma2007tuning}
S.~Varma, S.~B. Rempe, Tuning ion coordination architectures to enable
  selective partitioning, Biophys. J. 93~(4) (2007) 1093 -- 1099.

\bibitem{varma2011design}
S.~Varma, D.~M. Rogers, L.~R. Pratt, S.~B. Rempe, Design principles for {K$^+$}
  selectivity in membrane transport, J. Gen. Physiol. 137 (2011) 479--488.

\bibitem{rogers2011:probing}
D.~M. Rogers, S.~B. Rempe, Probing the thermodynamics of competitive ion
  binding using minimum energy structures, J. Phys. Chem. B 115~(29) (2011)
  9116--9129.

\bibitem{chaudhari2018strontium}
M.~I. Chaudhari, S.~B. Rempe, Strontium and barium in aqueous solution and a
  potassium channel binding site, J. Chem. Phys. 148 (2018) 222831.

\bibitem{varma2008valinomycin}
S.~Varma, D.~Sabo, S.~B. Rempe, {K$^+$/Na$^+$} selectivity in {K} channels and
  valinomycin: Over-coordination versus cavity-size constraints, J. Mol. Biol.
  376~(1) (2008) 13 -- 22.

\bibitem{Chaudhari:2014ga}
M.~I. Chaudhari, L.~R. Pratt, M.~E. Paulaitis, {Concentration dependence of the
  Flory-Huggins interaction parameter in aqueous solutions of capped PEO
  chains}, J. Chem. Phys. 141 (2014) 244908--5.

\bibitem{stevens2016:carboxylate}
M.~J. Stevens, S.~L. Rempe, Ion-specific effects in carboxylate binding sites,
  J. Phys. Chem. B 120~(49) (2016) 12519--12530.

\bibitem{rempe2001hydration}
S.~B. Rempe, L.~R. Pratt, The hydration number of {Na$^+$} in liquid water,
  Fluid Phase Equilib. 183 (2001) 121--132.

\bibitem{AsthagiriD:QuasB2}
D.~Asthagiri, L.~R. Pratt, {Quasi-chemical study of Be{$^{2+}$}(aq)
  speciation}, Chem. Phys. Lett. 371 (2003) 613--619.

\bibitem{rempe2004inner}
S.~B. Rempe, D.~Asthagiri, L.~R. Pratt, Inner shell definition and absolute
  hydration free energy of {K$^+$(aq)} on the basis of quasi-chemical theory
  and ab initio molecular dynamics, Phys. Chem. Chem. Phys. 6~(8) (2004)
  1966--1969.

\bibitem{asthagiri2004hydration}
D.~Asthagiri, L.~R. Pratt, M.~E. Paulaitis, S.~B. Rempe, Hydration structure
  and free energy of biomolecularly specific aqueous dications, including
  {Zn$^{2+}$} and first transition row metals, J. Am. Chem. Soc. 126~(4) (2004)
  1285--1289.

\bibitem{varma2008structural}
S.~Varma, S.~B. Rempe, Structural transitions in ion coordination driven by
  changes in competition for ligand binding, J. Am. Chem. Soc. 130~(46) (2008)
  15405--15419.

\bibitem{jiao2010:Ni}
D.~Jiao, K.~Leung, S.~B. Rempe, T.~M. Nenoff, First principles calculations of
  atomic nickel redox potentials and dimerization free energies: A study of
  metal nanoparticle growth, J. Chem. Theory Comput. 7~(2) (2010) 485--495.

\bibitem{Sabo:2013gs}
D.~Sabo, D.~Jiao, S.~Varma, L.~R. Pratt, S.~B. Rempe, {Case Study of
  Rb$^+$(aq), Quasi-Chemical Theory of Ion Hydration, and the No Split
  Occupancies Rule}, Ann. Rep. Prog. Chem, Sect. C (Phys. Chem.) 109 (2013)
  266--278.

\bibitem{chaudhari2015:Ba}
M.~I. Chaudhari, M.~Soniat, S.~B. Rempe, Octa-coordination and the aqueous
  {Ba$^{2+}$} ion, J. Phys. Chem. B 119~(28) (2015) 8746--8753.

\bibitem{chaudhari2016:carbonate}
M.~I. Chaudhari, J.~R. Nair, L.~R. Pratt, F.~A. Soto, P.~B. Balbuena, S.~B.
  Rempe, Scaling atomic partial charges of carbonate solvents for lithium ion
  solvation and diffusion, J. Chem. Theory Comput. 12~(12) (2016) 5709--5718.

\bibitem{chaudhari2018utility}
M.~I. Chaudhari, L.~R. Pratt, S.~B. Rempe, Utility of chemical computations in
  predicting solution free energies of metal ions, Mol. Sim. 44 (2018)
  110--116.

\bibitem{weber2012regularizing}
V.~Weber, D.~Asthagiri, Regularizing binding energy distributions and the
  hydration free energy of protein cytochrome {C} from all-atom simulations, J.
  Chem. Theory {\&} Comp. 8~(9) (2012) 3409--3415.

\bibitem{Tomar:2013dt}
D.~S. Tomar, D.~Asthagiri, V.~Weber, {Solvation Free Energy of the Peptide
  Group: Its Model Dependence and Implications for the Additive-Transfer
  Free-Energy Model of Protein Stability}, Biophys. J. 105 (2013) 1482--1490.

\bibitem{arshadi1970hydration}
M.~Arshadi, R.~Yamdagni, P.~Kebarle, Hydration of the halide negative ions in
  the gas phase. {II. C}omparison of hydration energies for the alkali positive
  and halide negative ions, J. Phys. Chem. 74 (1970) 1475--1482.

\bibitem{keesee1980properties}
R.~G. Keesee, N.~Lee, A.~Castleman~Jr, Properties of clusters in the gas phase:
  V. complexes of neutral molecules onto negative ions, J. Chem. Phys. 73
  (1980) 2195--2202.

\bibitem{hiraoka1988solvation}
K.~Hiraoka, S.~Mizuse, S.~Yamabe, Solvation of halide ions with {H}{$_2$}{O}
  and {CH$_3$CN} in the gas phase, J. Phys. Chem. 92 (1988) 3943--3952.

\bibitem{tissandier1998proton}
M.~D. Tissandier, K.~A. Cowen, W.~Y. Feng, E.~Gundlach, M.~H. Cohen, A.~D.
  Earhart, J.~V. Coe, T.~R. Tuttle, The proton's absolute aqueous enthalpy and
  gibbs free energy of solvation from cluster-ion solvation data, J. Phys.
  Chem. A 102 (1998) 7787--7794.

\bibitem{Tomasi:2005tc}
J.~Tomasi, B.~Mennucci, R.~Cammi, {Quantum Mechanical Continuum Solvation
  Models}, Chem. Rev. 105 (2005) 2999--3093, {N}ote Eq.~(74) and the following
  discussion regarding the factor of ``1/2" introduced there.

\bibitem{CPMD}
D.~Marx, J.~Hutter, Ab initio molecular dynamics: Theory and implementation,
  Modern Methods and Algorithms of Quantum Chemistry 1~(301-449) (2000) 141.

\bibitem{schlegel2002ab}
H.~B. Schlegel, S.~S. Iyengar, X.~Li, J.~M. Millam, G.~A. Voth, G.~E. Scuseria,
  M.~J. Frisch, Ab initio molecular dynamics: Propagating the density matrix
  with {G}aussian orbitals. iii. comparison with born--oppenheimer dynamics, J.
  Chem. Phys. 117 (2002) 8694--8704.

\bibitem{alvarez2009physiology}
F.~J. Alvarez-Leefmans, E.~Delpire, Physiology and Pathology of Chloride
  Transporters and Channels in the Nervous System: From Molecules to Diseases,
  Academic Press, 2009.

\bibitem{park2017structure}
E.~Park, E.~B. Campbell, R.~MacKinnon, Structure of a {CLC} chloride ion
  channel by cryo-electron microscopy, Nature 541 (2017) 500.

\bibitem{jentsch2002molecular}
T.~J. Jentsch, V.~Stein, F.~Weinreich, A.~a. Zdebik, Molecular structure and
  physiological function of chloride channels, Physiol. Rev. 82 (2002)
  503--568.

\bibitem{robertson2010design}
J.~L. Robertson, L.~Kolmakova-Partensky, C.~Miller, Design, function and
  structure of a monomeric {CLC} transporter, Nature 468 (2010) 844.

\bibitem{accardi2010clc}
A.~Accardi, A.~Picollo, {CLC} channels and transporters: Proteins with
  borderline personalities, Biochim. Biophys. Acta - Biomembranes 1798 (2010)
  1457--1464.

\bibitem{stockbridge2015crystal}
R.~B. Stockbridge, L.~Kolmakova-Partensky, T.~Shane, A.~Koide, S.~Koide,
  C.~Miller, S.~Newstead, Crystal structures of a double-barrelled fluoride ion
  channel, Nature 525 (2015) 548.

\bibitem{Sansom2002}
I.~Shrivastava, P.~Tieleman, P.~C. Biggin, M.~Sansom, {K$^+$} versus {Na$^+$}
  ions in a {K} channel selectivity filter: A simulation study, Biophys. J. 83
  (2002) 633--645.

\bibitem{Corry:2007}
M.~Thomas, D.~Jayatilaka, B.~Corry, The predominant role of coordination number
  in potassium channel selectivity, Biophys. J. 93~(8) (2007) 2635 -- 2643.

\bibitem{bostick2007selectivity}
D.~L. Bostick, C.~L. Brooks, Selectivity in {K$^+$} channels is due to
  topological control of the permeant ion's coordinated state, Proc. Natl.
  Acad. Sci., U.S.A. 104~(22) (2007) 9260--9265.

\bibitem{Fowler2008}
P.~Fowler, K.~Tai, M.~Sansom, The selectivity of {K$^+$} ion channels: Testing
  the hypotheses, Biophys. J. 95 (2008) 5062--5072.

\bibitem{bostick2009statistical}
D.~L. Bostick, C.~L. Brooks~III, Statistical determinants of selective ionic
  complexation: ions in solvent, transport proteins, and other “hosts”,
  Biophys. J. 96~(11) (2009) 4470--4492.

\bibitem{roux2011ion}
B.~Roux, S.~Bern{\`e}che, B.~Egwolf, B.~Lev, S.~Y. Noskov, C.~N. Rowley, H.~Yu,
  Ion selectivity in channels and transporters, J. Gen. Physiol. 137 (2011)
  415--426.

\bibitem{noskov2011importance}
S.~Y. Noskov, B.~Roux, Importance of hydration and dynamics on the selectivity
  of the {KcsA} and {NaK} channels, J. Gen. Physiol. 138 (2011) 651--651.

\bibitem{dixit2011thermodynamics}
P.~D. Dixit, D.~Asthagiri, Thermodynamics of ion selectivity in the {KcsA
  K$^+$} channel, J. Gen. Physiol. 137 (2011) 427--433.

\bibitem{Kim2011}
I.~Kim, T.~W. Allen, On the selective ion binding hypothesis for potassium
  channels, Proc. Natl. Acad. Sci., U.S.A. 108~(44) (2011) 17963--17968.

\bibitem{furini2011}
S.~Furini, C.~Domene, Selectivity and permeation of alkali metal ions in
  {K$^+$}-channels, J. Mol. Biol. 409~(5) (2011) 867 -- 878.

\bibitem{Andersen393}
O.~S. Andersen, Perspectives on: Ion selectivity, J. Gen. Physiol. 137 (2011)
  393--395.

\bibitem{Alam397}
A.~Alam, Y.~Jiang, Structural studies of ion selectivity in tetrameric cation
  channels, J. Gen. Physiol. 137~(5) (2011) 397--403.

\bibitem{Nimigean405}
C.~M. Nimigean, T.~W. Allen, Origins of ion selectivity in potassium channels
  from the perspective of channel block, J. Gen. Physiol. 137~(5) (2011)
  405--413.

\bibitem{roux2016}
D.~Medovoy, E.~Perozo, B.~Roux, Multi-ion free energy landscapes underscore the
  microscopic mechanism of ion selectivity in the {KcsA} channel, Biochim.
  Biophys. Acta Biomembr. 1858~(7, Part B) (2016) 1722 -- 1732.

\bibitem{deGroot2018}
W.~Kopec, D.~Kopfer, O.~Vickery, A.~S. Bondarenko, T.~L.~C. Jansen, B.~L.
  de~Groot, U.~Zachariae, Direct knock-on of desolvated ions governs strict ion
  selectivity in {K$^+$} channels, Nat. Chem. 10 (2018) 813--820.

\bibitem{gervasio2006exploring}
F.~L. Gervasio, M.~Parrinello, M.~Ceccarelli, M.~L. Klein, Exploring the gating
  mechanism in the {ClC} chloride channel via metadynamics, J. Mol. Bio. 361
  (2006) 390--398.

\bibitem{ko2010chloride}
Y.~J. Ko, W.~H. Jo, Chloride ion conduction without water coordination in the
  pore of {ClC} protein, J. Comp. Chem. 31 (2010) 603--611.

\bibitem{kuang2008transpath}
Z.~Kuang, A.~Liu, T.~L. Beck, Transpath: A computational method for locating
  ion transit pathways through membrane proteins, Proteins: Structure,
  Function, and Bioinformatics 71 (2008) 1349--1359.

\bibitem{yin2004ion}
J.~Yin, Z.~Kuang, U.~Mahankali, T.~L. Beck, Ion transit pathways and gating in
  {ClC} chloride channels, Proteins: Structure, Function, and Bioinformatics 57
  (2004) 414--421.

\bibitem{smith2011charge}
M.~Smith, H.~Lin, Charge delocalization upon chloride ion binding in {ClC}
  chloride ion channels/transporters, Chem. Phys. Lett. 502~(1-3) (2011)
  112--117.

\bibitem{chen2016free}
Z.~Chen, T.~L. Beck, Free energies of ion binding in the bacterial {ClC-Ec1}
  chloride transporter with implications for the transport mechanism and
  selectivity, J. Phys. Chem. B B 120 (2016) 3129--3139.

\bibitem{cheng2012molecular}
M.~H. Cheng, R.~D. Coalson, Molecular dynamics investigation of {Cl$^-$} and
  water transport through a eukaryotic {ClC} transporter, Biophys. J. 102
  (2012) 1363--1371.

\bibitem{Jordan:2007Notable}
P.~C. Jordan, {New and Notable: Tuning a potassium channel--the caress of the
  surroundings}, Biophys. J. 93~(4) (2007) 1091--1092.

\bibitem{ACR2019}
M.~I. Chaudhari, J.~Vanegas, A.~Muralidharan, L.~R. Pratt, S.~B. Rempe,
  Biomolecular hydration mimicry in ion permeation through membrane channels,
  Acc. Chem. Res.

\bibitem{chaudhari2017quasi}
M.~I. Chaudhari, S.~B. Rempe, L.~R. Pratt, Quasi-chemical theory of
  {F$^-$(aq)}: The “no split occupancies rule” revisited, J. Chem. Phys.
  147 (2017) 161728.

\bibitem{robertson2003molecular}
W.~H. Robertson, M.~A. Johnson, Molecular aspects of halide ion hydration: The
  cluster approach, Ann. Rev. Phys. Chem. 54 (2003) 173--213.

\bibitem{tobias2001surface}
D.~J. Tobias, P.~Jungwirth, M.~Parrinello, Surface solvation of halogen anions
  in water clusters: An ab initio molecular dynamics study of the
  {Cl$^-$(H$_2$O)$_6$} complex, J. Chem. Phys. 114 (2001) 7036--7044.

\bibitem{Marcus:1994ci}
Y.~Marcus, {A Simple Empirical Model Describing the Thermodynamics of Hydration
  of Ions of Widely Varying Charges, Sizes, and Shapes}, Biophys. Chem. 51
  (1994) 111--127.

\bibitem{Duignan:2017iha}
T.~T. Duignan, M.~D. Baer, G.~K. Schenter, C.~J. Mundy, {Real single ion
  solvation free energies with quantum mechanical simulation}, Chem. Sci. 8
  (2017) 6131--6140.

\bibitem{Soniat:2015}
M.~Soniat, D.~M. Rogers, S.~B. Rempe, {Dispersion- and Exchange-Corrected
  Density Functional Theory for Sodium Ion Hydration}, J. Chem. Theory and
  Comput. (2015) 150626121718001.

\bibitem{friedman1973thermodynamics}
H.~Friedman, C.~Krishnan, Thermodynamics of ionic hydration, in: Aqueous
  Solutions of Simple Electrolytes, Springer, 1973, pp. 1--118.

\bibitem{Merchant:2009fq}
S.~Merchant, D.~Asthagiri, {Thermodynamically dominant hydration structures of
  aqueous ions}, J. Chem. Phys. 130 (2009) 195102--11.

\bibitem{Willow:2017ij}
S.~Y. Willow, S.~S. Xantheas, {Molecular-Level Insight of the Effect of
  Hofmeister Anions on the Interfacial Surface Tension of a Model Protein}, J.
  Phys. Chem. Letts. 8~(7) (2017) 1574--1577.

\bibitem{g09}
M.~Frisch, G.~Trucks, H.~Schlegel, G.~Scuseria, M.~Robb, J.~Cheeseman,
  G.~Scalmani, V.~Barone, B.~Mennucci, G.~Petersson, Others, {G}aussian 09
  revision d. 01, 2009, {G}aussian Inc. Wallingford CT.

\bibitem{harvey1998flying}
S.~C. Harvey, R.~K.-Z. Tan, T.~E. Cheatham~{III}, The flying ice cube: Velocity
  rescaling in molecular dynamics leads to violation of energy equipartition,
  J. Comp. Chem. 19 (1998) 726--740.

\bibitem{Nose}
S.~Nos{\'e}, {A Molecular Dynamics Method for Simulations in the Canonical
  Ensemble}, Mol. Phys. 52 (1984) 255--268.

\bibitem{cp2k2005quickstep}
J.~VandeVondele, M.~Krack, F.~Mohamed, M.~Parrinello, T.~Chassaing, J.~Hutter,
  Quickstep: Fast and accurate density functional calculations using a mixed
  gaussian and plane waves approach, Comp. Phys. Comm. 167 (2005) 103--128.

\bibitem{perdew1996generalized}
J.~P. Perdew, K.~Burke, M.~Ernzerhof, Generalized gradient approximation made
  simple, Phys. Rev. Letts. 77 (1996) 3865.

\bibitem{goedecker1996separable}
S.~Goedecker, M.~Teter, J.~Hutter, Separable dual-space {G}aussian
  pseudopotentials, Phys. Rev. B 54 (1996) 1703.

\bibitem{lippert1999gaussian}
G.~Lippert, J.~Hutter, M.~Parrinello, The {G}aussian and augmented-plane-wave
  density functional method for ab initio molecular dynamics simulations, Theo.
  Chem. Accts. 103 (1999) 124--140.

\bibitem{vandevondele2007gaussian}
J.~VandeVondele, J.~Hutter, {G}aussian basis sets for accurate calculations on
  molecular systems in gas and condensed phases, J. Chem. Phys. 127 (2007)
  114105.

\bibitem{kresse1993ab}
G.~Kresse, J.~Hafner, Ab initio molecular dynamics for liquid metals, Phys.
  Rev.B 47 (1993) 558.

\bibitem{kresse1996efficient}
G.~Kresse, J.~Furthm{\"u}ller, Efficient iterative schemes for ab initio
  total-energy calculations using a plane-wave basis set, Phys. Rev. B 54
  (1996) 11169.

\bibitem{Gillan:2016ff}
M.~J. Gillan, D.~Alf{\`e}, A.~Michaelides, {Perspective: How good is DFT for
  water?}, J. Chem. Phys. 144 (2016) 130901--34.

\bibitem{Chaudhari:2017gs}
M.~I. Chaudhari, L.~R. Pratt, S.~B. Rempe, {Utility of chemical computations in
  predicting solution free energies of metal ions}, Mole. Sim. 492 (2017) 1--7.

\bibitem{Rempe:water}
S.~B. Rempe, T.~R. Mattsson, K.~Leung, On ``{The Complete Basis Set Limit}" and
  plane-wave methods in first-principles simulations of water, Phys. Chem.
  Chem. Phys 10 (2008) 4685--4687.

\end{thebibliography}
\end{document}


\renewcommand{\thepage}{S\arabic{page}}
\renewcommand{\thefigure}{S\arabic{figure}}

\title{Supporting Information: Quasi-Chemical Theory for Anion Hydration and Specific Ion Effects: Cl$^-$(aq) \emph{vs.} F$^-$(aq)}
\author{A. Muralidharan} 
\affiliation{Department of Chemical and Biomolecular Engineering, Tulane University, New Orleans, LA 70118}
\author{L. R. Pratt} 
\affiliation{Department of Chemical and Biomolecular Engineering, Tulane University, New Orleans, LA 70118}
\author{M. I. Chaudhari} 
\affiliation{Department of Nanobiology, Sandia National Laboratories,  Albuquerque, 87185, USA}
\author{S. B. Rempe*} 
\affiliation{Department of Nanobiology, Sandia National Laboratories,  Albuquerque, 87185, USA}

\maketitle
* Corresponding author: slrempe@sandia.gov\\
\section{Geometry optimizations:}
\begin{figure}[h]
\includegraphics[width=0.49\textwidth]{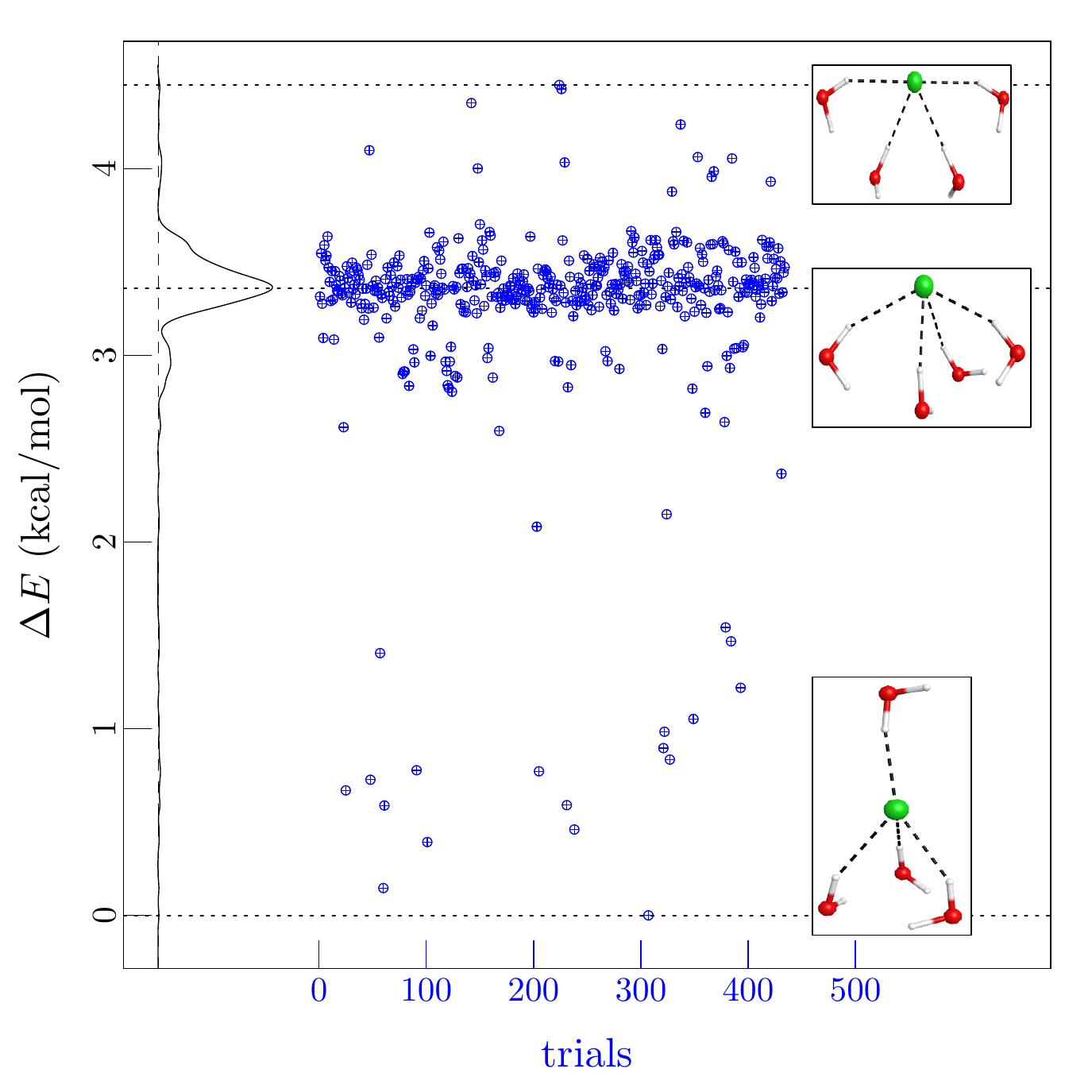}
\hfill
\includegraphics[width=0.49\textwidth]{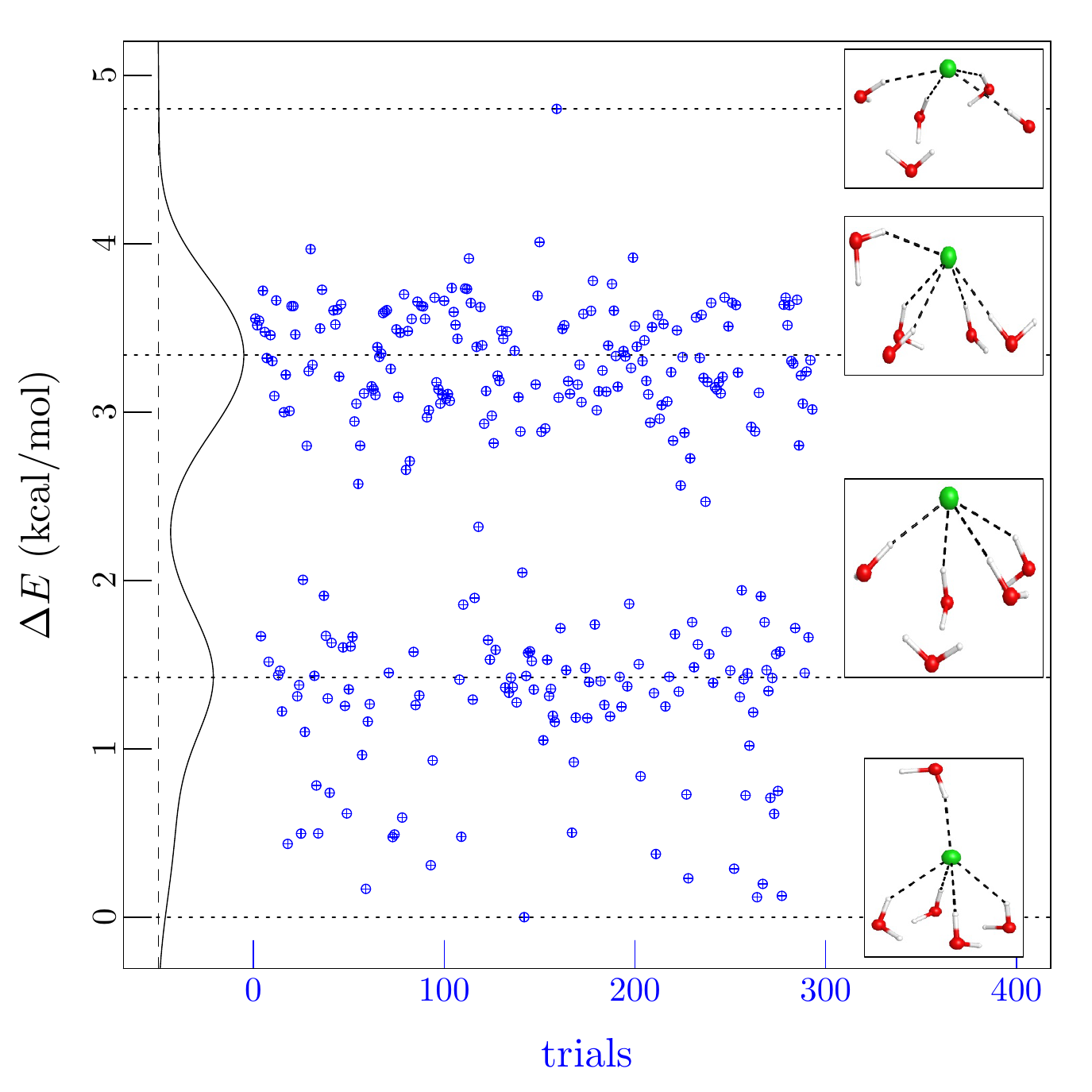}
\label{fig:Cl_sample_N5_2}
\caption{Blue dots: Electronic energy of optimized $\left(\mathrm{H}_2\mathrm{O}\right)_4\mathrm{Cl}^-$ (left) and $\left(\mathrm{H}_2\mathrm{O}\right)_5\mathrm{Cl}^-$ (right) clusters with initial configurations sampled from bulk phase AIMD. Black curve on the left shows the
distribution of those energies. The lowest optimum energy  (bottom inset) is about 4-5 kcal/mol lower in energy than
 the highest optimum energy (top inset). The harmonic approximation for estimation of cluster free energy is based on the lowest energy optimum. That structure is also used as a starting configuration for molecular dynamics with ADMP. }
\end{figure}
\FloatBarrier
\newpage
\section{Thermostat testing}
\begin{figure}[h]
\begin{center} \includegraphics[width=0.75\textwidth]{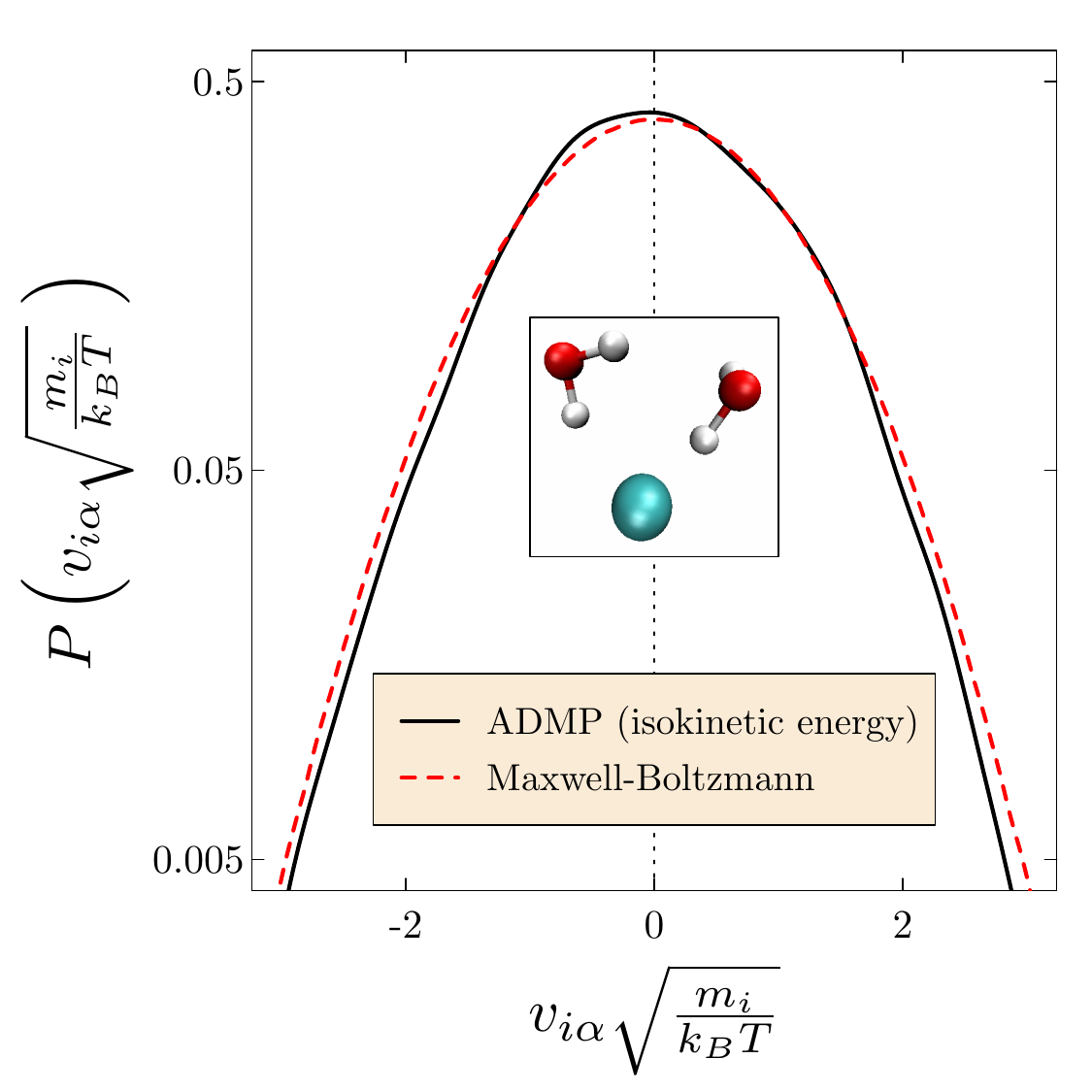} \end{center}
\caption{Testing the thermostat implemented with the ADMP approach\cite{schlegel2002ab} in the
Gaussian09 software package\cite{g09}
for the cluster $\left(\mathrm{H}_2\mathrm{O}\right)_2\mathrm{Cl}^-$ at
$T$=300~K. The UPBE1PBE density functional was utilized with the
aug-cc-pvdz basis set. Variable $\alpha = x, y, z$  indexes the Cartesian
components of atomic velocity. The observed distribution of atomic
velocities reasonably matches the Maxwell-Boltzmann distribution.} 
\label{fig:Admp_E} 
\end{figure}
\FloatBarrier
\newpage
\section{Comparison of cluster simulation methods:}
{Gaussian09 trajectory: The ADMP approach utilized the UPBE1PBE density functional with the aug-cc-pvdz basis set. The caveat of using Gaussian09 for dynamics is that the only available thermostat option is through the re-scaling of velocities. However, the velocity distribution obtained here is Maxwell-Boltzmann, as tested above (Fig \ref{fig:Admp_E}). For further testing, we use CP2K, as described below.\bigskip

    CP2K trajectory: CP2K allows molecular dynamics with the Nose-Hoover thermostat.\cite{Nose} The PBE density functional\cite{perdew1996generalized} was utilized with pseudopotentials proposed by Goedecker, Teter and Hutter (GTH\cite{goedecker1996separable}) in the hybrid Gaussian and plane wave scheme.\cite{lippert1999gaussian} Molecularly optimized basis sets denoted as DZVP-MOLOPT-GTH\cite{vandevondele2007gaussian} were obtained from the CP2K website. The trajectory obtained is then analysed in Gaussian09 using the UPBE1PBE density functional with the aug-cc-pvdz basis set.}
\begin{figure}[h]
\includegraphics[width=0.49\textwidth]{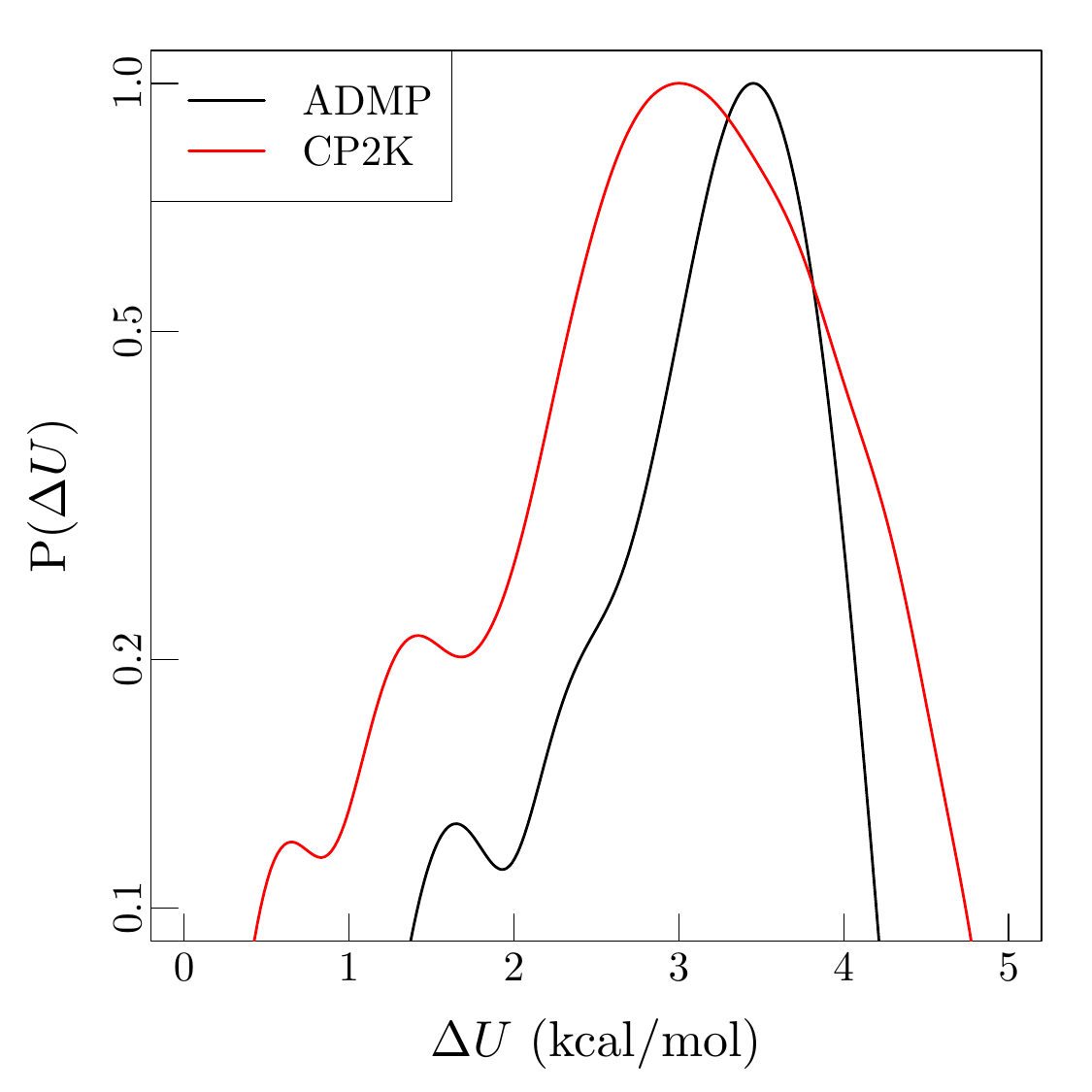}
\hfill
\includegraphics[width=0.49\textwidth]{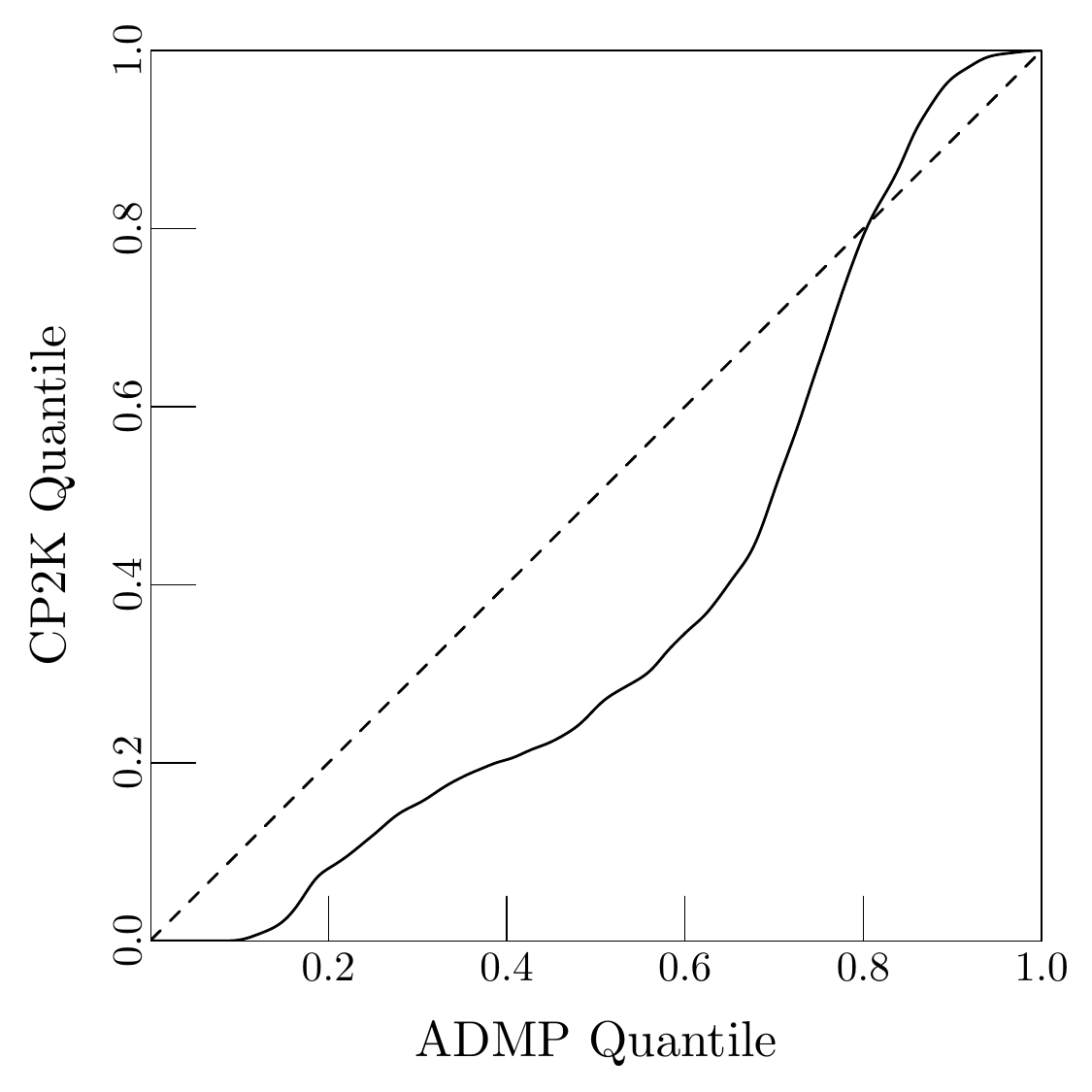}
\label{Q-Q}
\caption{Left: Comparison of $\Delta U$ samples for $\left(\mathrm{H}_2\mathrm{O}\right)_2\mathrm{Cl}^-$ obtained using  the ADMP approach\cite{schlegel2002ab} (black) in the Gaussian09 software package 
with the mixed Gaussian and plane wave approach (red) in the CP2K package.\cite{cp2k2005quickstep} $\Delta U = E\left(\gamma_n\sigma \right) - E\left(\gamma_{n-1}\sigma \right) - E\left(\gamma\sigma \right) + E(\sigma)~$, defined in the main text.
Right: The standard Q-Q plot for the distribution shown in the left panel.}
\end{figure}

\bibliographystyle{achemso}
\FloatBarrier
\bibliography{Bibliography.bib}